\documentclass[11pt,a4paper]{iopart}

\usepackage{graphicx,color}
\usepackage{cite}
\usepackage[colorlinks,linkcolor=blue,urlcolor=blue,citecolor=blue]{hyperref}
\usepackage{xcolor}

\usepackage{iopams}

\expandafter\let\csname equation*\endcsname\relax

\expandafter\let\csname endequation*\endcsname\relax 

\usepackage{amsmath}
\usepackage{amssymb}

\makeatletter
\def\@mkboth#1#2{}
\newlength\appendixwidth
\newcommand{\patchl@section}{%
  \settowidth{\appendixwidth}{\textbf{Appendix }}%
  \addtolength{\appendixwidth}{1.5em}%
  \patchcmd{\l@section}{1.5em}{\appendixwidth}{}{\ddt}%
}
\makeatother

\definecolor{orange}{RGB}{20, 150, 60}

\newcommand{\eqnref}[1]{\eref{#1}}%
\newcommand{\figref}[1]{Fig.~\ref{#1}} %
\newcommand{\secref}[1]{Sec.~\ref{#1}}%
\newcommand{\Secref}[1]{Section~\ref{#1}}%

\begin{document}
\title[Run and tumble particle in $U(x)=\alpha |x|$ potential with an absorbing boundary]
{Survival probability and position distribution of
a run and tumble particle in $U(x)=\alpha |x|$ potential with an
absorbing boundary}
\author{Sujit Kumar Nath$^{1,2,3,*}$, Sanjib Sabhapandit$^{4,\dag}$}
\address{$^1$ Aix Marseille Univ, Université de Toulon, CNRS, CPT (UMR 7332), Turing
Center for Living Systems, Marseille, France\\
$^2$ Division of Cardiovascular Sciences, Faculty of Biology, Medicine and Health,  The University of Manchester, M13 9PT, Manchester, United Kingdom \\
$^3$ School of Computing and Faculty of Biological
Sciences, University of Leeds, LS2 9JT, Leeds, United Kingdom \\
$^{4}$ Raman Research Institute, Bangalore 560080, India}
\ead{$^{*}$sujit-kumar.nath@univ-amu.fr, $^\dag$sanjib@rri.res.in}

\begin{abstract}
We study the late time exponential decay of the survival probability $S_\pm(t,a|x_0)\sim e^{-\theta(a)t}$, of a one-dimensional run and tumble particle starting from $x_0<a$ with an initial orientation $\sigma(0)=\pm 1$, under a confining potential $U(x)=\alpha|x|$ with an absorbing boundary at $x=a>0$.
We find that the decay rate $\theta(a)$ of the survival probability has strong dependence
on the location $a$ of the absorbing boundary, which undergoes a freezing transition
at a critical value $a=a_c=(v_0-\alpha)\sqrt{v_0^2-\alpha^2}/(2\alpha\gamma)$, where $v_0>\alpha$ is the self-propulsion speed and $\gamma$ is the tumbling rate of the particle. For $a>a_c$, the value of $\theta(a)$ increases
monotonically from zero, as $a$ decreases from infinity, till it attains the maximum value $\theta(a_c)$ at $a=a_c$.
For $0<a<a_c$, the value of $\theta(a)$ freezes to the value $\theta(a)=\theta(a_c)$.
We also obtain the propagator with the absorbing boundary condition at $x=a$. Our analytical results are supported by numerical simulations.
\end{abstract}

\section{Introduction}
The applications of
target-search problem can be found across disciplines, such as
physics~\cite{evans2011diffusion,kusmierz2014first,
evans2020stochastic,chupeau2020optimizing,evans2018run,chupeau2015cover,
eliazar2023hazard,cherayil2022effects,PhysRevLett.123.250603},
ecology~\cite{benichou2011intermittent,viswanathan2011physics},
cellular processes in biology~\cite{sirovich2011spiking,Gomez_2020,
PhysRevX.6.041037,shin2019target},
computer science~\cite{vergassola2007infotaxis} etc. In this context, the time at which  the searcher reaches the target for the first time,
known as the first passage
time (FPT), is a relevant quantity. The  probability density function (PDF) $F(t,a|x_0)$ of the FPT to a target located at $x=a$ and its cumulative distribution
$S(t,a|x_0)=\int_t^\infty F(t',a|x_0)dt'$, known as the survival probability 
with an absorbing boundary at $x=a$, starting from $x_0$, are two of the most
well-studied quantities in random processes.
Animals, bacteria, and several other species forage for their foods,
mates or shelters in a random fashion~\cite{berg1993random}.
Therefore, the natural questions of interest are ``How long does a forager need to encounter its target?'',
``How does this hitting time depend on the location of the target?'', etc.
Also, it is quite naturally observed that a forager always searches for its
target starting from a home-location and remains confined in a territory
around it. Even for a few animal species, such as the desert ants, being confined
in a territory around their nests, while searching for food, is so crucial
that any violation of this condition may be life threatening for them~\cite{wehner2020desert}.
Therefore, there is always a pull on the forager toward its home-territory,
which influences the forager not to go too far from its home while executing
the search process. One simple way to model such
random movements executed by a forager, is to consider
the run and tumble dynamics, where the forager chooses its direction of movement randomly
and moves in that direction at a constant speed~\cite{malakar2018steady,santra2020run,santra2023long,radice2021one}.
The pull toward its home can be modelled by a confining potential having its minimum at the
location of the forager's nest or starting location~\cite{smith2022exact,sevilla2019stationary,smith2022nonequilibrium,santra2021brownian,goerlich2023experimental,mercado2020intermittent,mercado2022reducing}.

The study of first passage problems was started since the time of
Sch\"ordinger and Smoluchowski \cite{schrodinger1915theorie,smoluchowski1915}.
Later Kramers studied
the escape rate of a Brownian particle over a potential barrier under the
influence of thermal fluctuations by computing the flux across the barrier, and found that it is well approximated by the van't Hoff-Arrhenius form at low temperature and high barrier height limit~\cite{kramers1940brownian,farkas1927keimbildungsgeschwindigkeit,
kaischew1934kinetischen}. It has also been found that the other alternative approaches
using FPT and survival probability can be useful to study such escape rates~\cite{chandrasekhar1943stochastic,chandrasekhar1943dynamical,klein1952mean,redner2001guide,pontryagin1989statistical}.
Since then the study of FPT and survival probability has been extensively developed.
Although the mean first passage time (MFPT) has been extensively studied in
various contexts~\cite{gueneau2024optimal,rupprecht2016optimal,angelani2023one}, the full distribution of the FPT, or equivalently, the survival probability carries much more information. However, the computation of the FPT distribution and the survival probability are much harder compared to the MFPT.
\cite{riskenFP,majumdar1999persistence,redner2001guide,majumdar2007brownian,
bray2013persistence}.

It is evident that the distribution of the FPT
or the survival probability must depend on the distance between the location of the absorbing barrier and the starting point. Therefore,
understanding the dependence of the FPT and survival probability on the position of the
absorbing barrier is of immense importance to have a clear understanding of
the random search problems. For a Brownian particle in a `sufficiently' confining potential, generically, $S(t,a|0)\sim e^{-\theta(a)t}$, where $x=0$ being the starting point of the particle as well as the minima of the potential. Recently, it has been found that for potentials having the property $U(x)\sim |x|$ when $x\to-\infty$, the decay rate
$\theta(a)$ undergoes a freezing
transition depending on the position of the barrier $a$~\cite{PhysRevLett.125.200601}.
Above a critical value $a=a_c$, $\theta(a)$ decreases with increasing value of $a$.
However, $\theta(a)$ freezes to a constant value below a critical value $a_c$.
Other than the decay rate, freezing transitions of other parameters
such as the strength of the confining potential which minimizes the MFPT, depending on
the position of the absorbing boundary, has also been studied for confined Brownian particles~\cite{mercado2022freezing}.
Similar transitions in various parameter values (such as resetting rate, potential exponent, etc.) which minimize MFPT, have also been studied in models with stochastic resetting  and reflecting boundaries~\cite{ahmad2022first,singh2020resetting,ray2021resetting,ahmad2019first,capala2023optimization}.
Therefore, a natural question arises whether such freezing transitions exist for other
random processes, such as for a run and tumble particle (RTP) which closely resembles
the bacteria or animal movements.

The steady state, relaxation time, extreme values, first passage properties
to the origin (minima of the potential) from a position away from the
origin have been previously studied for an RTP in various types of confining potential~\cite{dhar2019run,le2021stationary,mori2021distribution,gueneau2024optimal,de2021survival}. In particular, for a confining potential of the form $U(x)= \alpha |x|^\nu$, without any absorbing barrier, the steady state distribution of the position is supported within a finite interval $[-(v_0/(\alpha\nu))^{1/(\nu-1)}, (v_0/(\alpha\nu))^{1/(\nu-1)}]$ for $\nu>1$~\cite{dhar2019run}, where $v_0$ denotes the self-propulsion speed.
Therefore, in this case an RTP starting at the origin cannot reach a target position outside this interval. On the other hand for $\nu<1$, the stationary distribution tends to a delta function at the origin, i.e., $P_{\text{st}}(x)=\delta(x)$~\cite{dhar2019run}.
In the marginal case $\nu=1$, with $v_0>\alpha$, the stationary state is supported over the entire line $-\infty<x<\infty$~\cite{dhar2019run}.
In this article we study the first passage properties to a location away from the origin for a one-dimensional RTP in a confining potential $U(x)=\alpha |x|$.
This is the same potential where the freezing transition of the decay rate $\theta(a)$ was observed for a Brownian particle~\cite{PhysRevLett.125.200601}.
A natural question is whether such freezing transition also occurs for an active particle in the same potential.

Here we calculate the late time behavior of the survival probability for an RTP in the potential $U(x)=\alpha |x|$ for $-\infty<x<a$, with an absorbing boundary at $x=a>0$.
We find that starting at the origin with the orientation $\sigma(0)=\pm1$, the survival probability decays exponentially, $S_\pm(t,a|0)\sim e^{-\theta(a)t}$, at long times.
The location dependent decay exponent $\theta(a)$ indeed undergoes a freezing transition, similar to the one observed in the passive case~\cite{PhysRevLett.125.200601}.
There is a critical value of the location $a=a_c=(v_0-\alpha)\sqrt{v_0^2-\alpha^2}/(2\alpha\gamma)$ such that $\theta(a)$ is constant for $0<a<a_c$ and decreases monotonically with increasing $a>a_c$. We also obtain the propagator in the presence of the potential with an absorbing boundary at $x=a$.

The paper is organized as follows. In \secref{model} we define the model and obtain the survival probability using a backward Fokker-Planck equation approach. In \secref{mfptDecayComp} we present a comparison between the inverse of the MFPT and the  decay
rate $\theta(a)$. In \secref{RTP-propagator}, we obtain the propagator for the position of the particle in the presence of the absorbing barrier. Finally, we summarize our findings in \secref{conclusion}.
We present an alternative derivation of the survival probability from the propagator in \ref{survFwdFP}. Some details of the calculations are relegated to \ref{appnSur}.

\section{Run and Tumble particle in \texorpdfstring{$U(x)=\alpha |x|$}{U(x)=alpha |x|} potential}
\label{model}

We consider an RTP in one dimension, subjected to an external potential $U(x)=\alpha |x|$. The equation of motion reads
\begin{equation}
\frac{dx}{dt} = -\, \alpha \, \mathrm{sgn}(x) + v_0 \, \sigma (t),
\label{eq102}
\end{equation}
where $\sigma(t)$ is a dichotomous telegraphic noise that alternates between $\pm 1$ intermittently with a constant Poissonian rate $\gamma$. The self-propulsion speed $v_0$ is a constant.
There is an absorbing boundary at a position $x=a > 0$. Starting at a position $x_0<a$, we want to obtain the PDF of finding the RTP at position $x$ at time $t$, and the probability that the  particle survives (i.e., it does not cross the position $a$) up to time $t$.
\begin{figure}
\begin{center}
\includegraphics[scale=0.3]{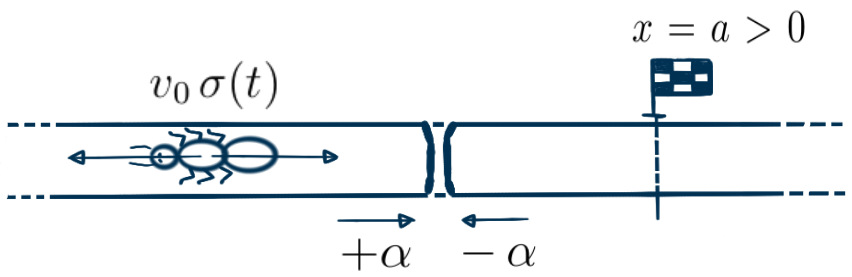} \\[0.6cm]
\includegraphics[scale=0.55]{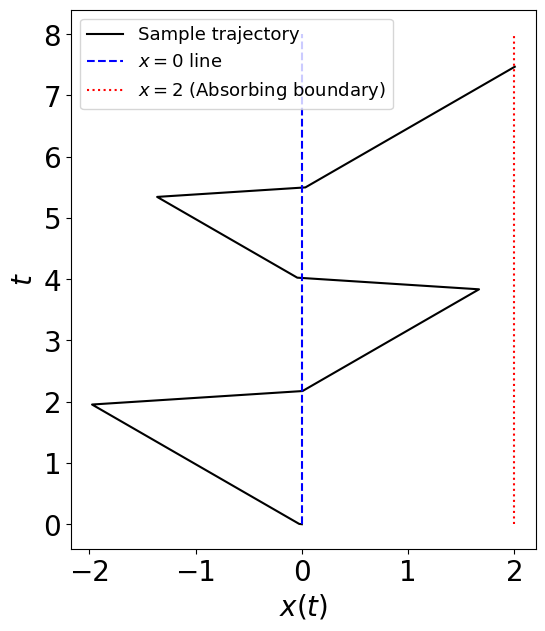}
\caption{The panel above is a cartoonist's bug-on-a-conveyor-belt representation (top view) of the model. The left and right conveyor belts are running with velocities $+\alpha$ and $-\alpha$ respectively, in the opposite directions, resembling the attractive potential $U(x)=\alpha|x|$. The bug on the conveyor belt is the run and tumble random walker running with a randomly switching velocity $v_0 \sigma(t)$ without any slipping. The position of the flag on the ground is the location of the absorbing boundary. The panel below shows a sample path of the random walker.}
\label{figModelCartoon}
\end{center}
\end{figure}
Note that for any $\alpha > v_0$, the net velocity is always toward the origin, and hence for any $x_0 <a$, the particle moves toward the origin and arrives the origin at a finite time, and subsequently sits at the origin forever.  Here, we consider the more interesting case of $\alpha <v_0$. A cartoonist's representation and a sample path of the model are shown in \figref{figModelCartoon}.
As our primary goal in this paper is to investigate
the existence of a freezing transition of the survival probability
depending on the position of the absorbing boundary, we present the
detailed study of the survival probability in the following sections,
postponing the study of the position distribution in the \secref{RTP-propagator}.
\subsection{Survival probability}
\label{spMain}
The survival probability can be most conveniently derived using a backward Fokker-Planck approach. To this end, let
$S_\pm (t,a|x_0)$  be the survival probability, i.e.,  the probability that an RTP starting from the position $x_0<a$, with $\sigma(0)=\pm 1$ respectively, does not cross the location $a>0$ (hence survives) up to time $t$.
Starting with \eqnref{eq102}, and analyzing the starting move within an initial infinitesimal time interval $[0,dt]$, we obtain the backward Fokker-Planck equation for $S_\pm (t,a|x_0)$ as~\cite{riskenFP}
\begin{align}
\label{survBkdFP1}
\frac{\partial S_+}{\partial t} &= \bigl[v_0-\alpha\,\mathrm{sgn}(x_0)\bigr] \frac{\partial S_+}{\partial x_0} -\gamma S_+ +\gamma S_-, \\
\label{survBkdFP2}
\frac{\partial S_-}{\partial t} &= -\big[v_0+\alpha\,\mathrm{sgn}(x_0)\bigr]  \frac{\partial S_-}{\partial x_0} -\gamma S_- +\gamma S_+,
\end{align}
with the initial condition $S_\pm (0,a|x_0)=1$.
A particle starting at the absorbing boundary with positive velocity, i.e., $\sigma(0)=1$, crosses the boundary immediately, implying the boundary condition
$S_+(t,a|a)=0$. On the other hand, if the particle starts from the boundary with a negative velocity, i.e., $\sigma(0)=-1$, it does not get absorbed immediately. Therefore, the boundary condition $S_-(t,a|a)$ is not specified. A particle starting from
$-\infty$, cannot cross the level $a$ at any finite time, and hence, always survives, yielding the boundary condition
$S_\pm (t,a|-\infty)=1$.

To proceed, it is convenient to take Laplace transform of Eqs. \eref{survBkdFP1} and \eref{survBkdFP2}, with
 respect to $t$, which gives
\begin{align}
\label{eq105}
&\bigl[v_0-\alpha\,\mathrm{sgn}(x_0)\bigr] \tilde{S}'_+ - (\gamma+s) \tilde{S}_+ +\gamma \tilde{S}_-=-1,\\
\label{eq106}
&\bigl[v_0+\alpha\,\mathrm{sgn}(x_0)\bigr] \tilde{S}'_- + (\gamma+s) \tilde{S}_- -\gamma \tilde{S}_+=1,
\end{align}
where we have already used the initial condition.
The boundary conditions become $\tilde{S}_\pm(s,a|-\infty)=1/s$ and $\tilde{S}_+(s,a|a)=0$. 
Solving \eqnref{eq105} and \eqnref{eq106} with these boundary conditions we get [see \secref{calsSurv}]

\begin{equation}
\label{eq144.aMain}
\tilde{S}_\pm(s,a|x_0)=\frac{1}{s}\Bigl[1-\tilde{F}_\pm (s,a|x_0) \Bigr],
\end{equation}
where
\begin{equation}
\tilde{F}_\pm(s,a|x_0)=\frac{p}{\mathbb{D}}\, e^{-[\alpha (\gamma+s)(a-|x_0|)+pa]/(v_0^2-\alpha^2)}\, \tilde{\psi}_\pm(s|x_0),
\label{eqFpmSX0}
\end{equation}
with $p$ and $\mathbb{D}$ are given by 
\begin{align}
\label{eqp}
p &=\sqrt{v_0^2(\gamma+s)^2-\gamma^2(v_0^2-\alpha^2)},\\
 \label{eqD}
\mathbb{D} &=v_0\bigl[p-\alpha (\gamma+s) \bigr]-\alpha \bigl[p-v_0(\gamma+s) \bigr] e^{-2 p a/(v_0^2-\alpha^2)}.
\end{align}
The functions 
$\tilde{\psi}_\pm(s|x_0)$ in \eqnref{eqFpmSX0} are  given by, \\
{\bf (1) for $0 < x_0 \le a$:}
\begin{align} 
\tilde{\psi}_+(s|x_0)=&\frac{1}{p}\,
\Bigl\{
v_0\bigl[p-\alpha (\gamma+s) \bigr]e^{px_0/(v_0^2-\alpha^2)}-\alpha \bigl[p-v_0(\gamma+s) \bigr] \,e^{- p x_0/(v_0^2-\alpha^2)}
\Bigr\},\\
\nonumber
\tilde{\psi}_-(s|x_0)=&
\left[\frac{v_0(\gamma+s)-p}{p\gamma(v_0+\alpha)}\right]
\Bigl\{
v_0\bigl[p-\alpha (\gamma+s) \bigr]\,e^{px_0/(v_0^2-\alpha^2)}\\ 
&\qquad\qquad\qquad\quad+\alpha \bigl[p+v_0(\gamma+s) \bigr] e^{- p x_0/(v_0^2-\alpha^2)}
\Bigr\},
\end{align}
{\bf (2) for $x_0<0$:}
\begin{align}
\label{eq:psiPBackward}
\tilde{\psi}_+(s|x_0)&=(v_0-\alpha)\, e^{px_0/(v_0^2-\alpha^2)}\\
\label{eq:psiNBackward}
\tilde{\psi}_-(s|x_0)&=\frac{1}{\gamma}\,\big[v_0(\gamma+s)-p\bigr]\, e^{px_0/(v_0^2-\alpha^2)}.
\end{align}
Similar expressions for the Laplace transform of the survival probability, as
in \eqnref{eq144.aMain}, have previously been presented in~\cite{mori2022time}. However, 
the detailed study of the survival probability in time domain, including the freezing transition discussed in our paper, was not carried out.
Evidently $\tilde{S}_\pm(s,a|x_0)$ can also be obtained by integrating the final
position of the propagator with an absorbing boundary condition at $x=a$, as shown in \ref{survFwdFP}. We obtain the propagator with the absorbing boundary condition by solving the forward Fokker-Planck
equation in \Secref{RTP-propagator}.

\begin{figure}
\begin{center}
\includegraphics[scale=0.6]{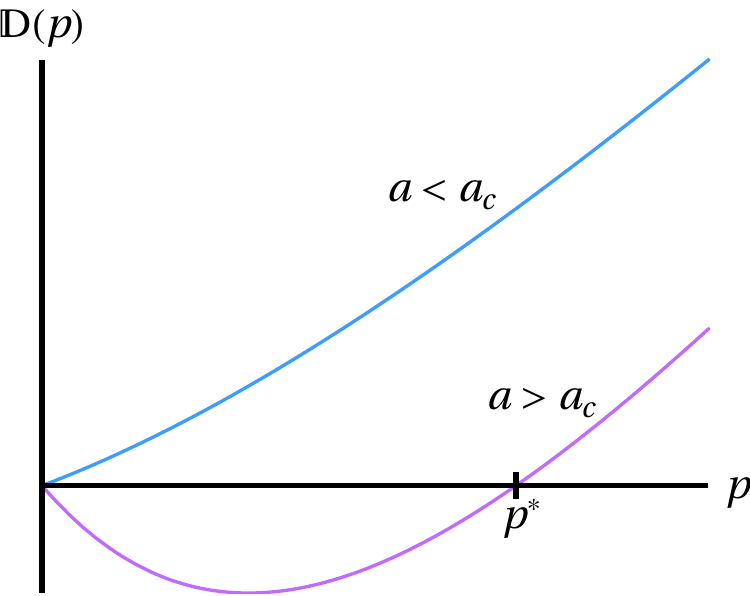}
\caption{\label{active-pstar-fig}  Graphical representation of the solution of the equation $\mathbb{D}=0$, where $\mathbb{D}$ is given by \eqnref{eqn168}. The trivial solution is $p=0$. There exists a non-trivial solution $p^*\in (0,\alpha\gamma)$  for $a>a_c$.
}
\end{center}
\end{figure}

In the time domain, the survival probability is given by the
Bromwich integral
\begin{equation}
S_\pm(t,a|x_0)=\frac{1}{2\pi i}\int_{0^+-i\infty}^{0^++i\infty} \tilde{S}_\pm(s,a|x_0) \, e^{st}\, ds,
\label{eq163}
\end{equation}
where $\tilde{S}_\pm(s,a|x_0)$ are given by \eqnref{eq144.aMain}. For simplicity, now we consider the initial position $x_0=0$, for which
\begin{equation}
\tilde{F}_\pm (s,a|0)= \frac{p}{\mathbb{D}} \, e^{-[p+\alpha(\gamma+s)]a/(v_0^2-\alpha^2)}\, \tilde{\psi}_\pm(s),
\end{equation}
with
\begin{align}
\tilde{\psi}_+ (s) &= (v_0-\alpha),\\
\tilde{\psi}_- (s) &= \frac{1}{\gamma} \, [v_0(\gamma+s)-p],
\end{align}
where $\mathbb{D}$ is given by \eqnref{eqD} and $p$ is given by \eqnref{eqp}. For $a\to \infty$, we have $\tilde{F}_\pm(s,a|0)\to 0$. Therefore, $\tilde{S}_\pm(s,a|0)\to 1/s$ has a simple pole at $s=0$, and consequently, $S_\pm(t,a|0)=1$. On the other hand, for any finite $a$, we get $\tilde{F}_\pm(s\to0,a|0)=1+O(s)$. Therefore, $s=0$ is no longer a pole and the most dominant contribution comes from the singularity closer to $s=0$. Note that $p=0$ corresponds to  the pair of branch-points at
\begin{equation}
\label{sbpm}
s_\mathrm{b}^\pm=\gamma \left[-1\pm\sqrt{1-(\alpha/v_0)^2} \right].
\end{equation}
On the other hand, if $\mathbb{D}=0$ has a solution for real $s$ within
$(s_\mathrm{b}^+, 0)$,
then it becomes the most dominant singularity (a simple pole). Expressing $\mathbb{D}$ completely in terms of $p$, we get
\begin{equation}
\mathbb{D}(p)= v_0\left[p-(\alpha/v_0)\sqrt{p^2 +\gamma^2 (v_0^2-\alpha^2)}  \right]
-\alpha \left[p-\sqrt{p^2 +\gamma^2 (v_0^2-\alpha^2)}  \right] e^{-2 p a/(v_0^2-\alpha^2)} .
\label{eqn168}
\end{equation}
Note that although $\mathbb{D}(p=0)=0$, it does not correspond to a pole as $\tilde{F}_\pm (s,a|0) \propto p/\mathbb{D}$.
It is easy to check that $\mathbb{D}(p) = (v_0-\alpha)p + \mathcal{O}(1/p)$ as $p\to \infty$. On the other hand, for small $p$, we get
\begin{equation}
\label{dpAleAc}
\mathbb{D}(p)=\frac{2\alpha\gamma \,(a_c-a)}{\sqrt{v_0^2-\alpha^2}}\,  p + \mathcal{O}(p^2),
\end{equation}
with
\begin{equation}
a_c = \frac{(v_0-\alpha) \sqrt{v_0^2-\alpha^2}}{2\alpha \gamma}.
\end{equation}
Therefore,  $\mathbb{D}(p)$ for positive $p$, stays always positive for any $a < a_c$, and consequently, the contribution comes from the branch-point $s_\mathrm{b}$.
On the other hand, for $a >a_c$, $\mathbb{D}(p) <0 $ for small $p$,   then  at some value $p=p^*$, $\mathbb{D}(p^*)=0$, and $\mathbb{D}(p) >0$ for $p > p^*$ [see \figref{active-pstar-fig}]. Therefore, for $a >a_c$, $\tilde{F}_\pm (s,a|0)$ has a pole at
\begin{equation}
s^*=-\gamma + \frac{1}{v_0} \sqrt{p^{*2} + \gamma^2(v_0^2-\alpha^2)}.
\label{sstarPstar}
\end{equation}
Note that $p^*\in (0,\alpha\gamma)$, with $p^*\to 0$, as $a \to a_c^+$ and $p^*\to \alpha\gamma$ as $a\to \infty$. Consequently, $s^*\in (s_\mathrm{b}^+,0)$ with  $s^*\to s_\mathrm{b}^+$ as $a\to a_c^+$ and $s^*\to 0$ as $a\to \infty$.
Therefore, , we have
\begin{align}
S_\pm (t,a|0) =
\left\{
\begin{array}{ll}
Q_\pm(t,a) \, e^{-|s_\mathrm{b}^+|\, t} , &\text{for $a\le a_c$}\\[2mm]
R_\pm(a)  \, e^{-|s^*| \, t}  + Q_\pm(t,a) \, e^{-|s_\mathrm{b}^+|\, t}  &\text{for $a> a_c$}.
\end{array}
\right.
\end{align}
where $R_\pm(a)=\lim_{s\to s^*} (s-s^*) \tilde{S}_\pm (s,a|0)$ arise from the residue at $s^*$ and are given by
\begin{equation}
R_\pm(a)=\frac{p^*\, e^{-[p^*+\alpha(\gamma+s^*)]a/(v_0^2-\alpha^2)}\, \tilde{\psi}_\pm(s^*)}{(-s^*)\mathbb{D}'(p^*)\, p'(s^*)}.
\label{rpmA}
\end{equation}
The prefactor $Q_\pm(t,a)$ comes from the integral around the branch-cut from $s_\mathrm{b}^-$ to $s_\mathrm{b}^+$, which we determine below.
\subsection{Leading order decay of the survival probability}
Although obtaining the exact analytical expressions for the survival probabilities
in the time domain, by integrating \eqnref{eq163}, seems  difficult, one can still obtain their leading order behavior
at large time. \\ \\
\noindent
{\bf Case-1 ($a>a_c$):}
As discussed previously, after \eqnref{eqn168}, when the absorbing
boundary is at $a>a_c$, the
decay of the survival probability at large time is dominated by the
pole $s^*(a)$ in the interval $(s_b^+,0)$, and the scaling of the decay
is given by $R_{\pm}(a)e^{-|s^*(a)|t}$.
\\ \\
\noindent
{\bf Case-2 ($a=a_c$):}
In the case of $a=a_c$, we have the leading order for $\mathbb{D}(p)$
as
\begin{equation}
\mathbb{D}(p)=
\frac{p^2 \sqrt{v_0^2-\alpha ^2}}{2 \alpha  \gamma }+O\left(p^3\right),
\label{dPleadAeqAC}
\end{equation}
and the pole disappears. Hence, the decay is now dominated
by the integral of $\tilde{S}_{\pm}(s,a|0)e^{st}$ around the branch cut from
$s_b^-$ to $s_b^+$, and therefore, dominated by $Q_{\pm}(t,a_c)e^{-|s_b^+|t}$.
We now estimate the leading order dependence of $Q_{\pm}(t,a_c)$, on $t$, for
$a=a_c$. Taking the series expansion of $\tilde{S}_\pm(s,a|0)$ near $s_b^+$ in the right hand side of \eqnref{eq163} and integrating [see \ref{appnSur} for the details]
we obtain the leading order $t$-dependence of $Q_{\pm}(t,a_c)$ as
\begin{equation}
Q_{\pm}(t,a_c)\sim f_{\pm}(v_0,\alpha,\gamma,a_c)~t^{-1/2},
\label{sxtscaleAeqAC}
\end{equation}
where $f_{\pm}(v_0,\alpha,\gamma,a_c)$ are functions of $v_0,\alpha,\gamma$, and $a_c$, given by
\begin{equation}
\hspace{-2cm}
f_{+}(v_0,\alpha,\gamma,a_c)=
   \frac{\alpha\sqrt{2v_0\sqrt{v_0^2-\alpha^2}}}
   {\sqrt{\pi \gamma}(v_0+\alpha)(v_0-\sqrt{v_0^2-\alpha^2})}~\exp\left[-\frac{a_c\alpha\gamma}{v_0\sqrt{v_0^2-\alpha^2}}\right]
\label{sxtscaleAeqACCoeffp}
\end{equation}
\begin{equation}
\hspace{-2cm}
f_{-}(v_0,\alpha,\gamma,a_c)=
 \frac{\alpha\sqrt{2v_0\sqrt{v_0^2-\alpha^2}}}
   {\sqrt{\pi\gamma}\sqrt{v_0^2-\alpha^2}(v_0-\sqrt{v_0^2-\alpha^2})}~\exp\left[-\frac{a_c\alpha\gamma}{v_0\sqrt{v_0^2-\alpha^2}}\right]
\label{sxtscaleAeqACCoeffn}
\end{equation}
\noindent
{\bf Case-3 ($a<a_c$):}
Similarly, when $a<a_c$, the decay is dominated
by the integral of $\tilde{S}_{\pm}(s,a|0)e^{st}$ in \eqnref{eq163} around the branch cut from
$s_b^-$ to $s_b^+$ as there is no pole (can be seen from the series
expansion of $\mathbb{D}(p)$ in \eqnref{dpAleAc}),
which is given by $Q_{\pm}(t,a)e^{-|s_b^+|t}$.
Taking the series expansion of $\tilde{S}_{\pm}(s,a|0)$
near $s_b^+$ in the right hand side of \eqnref{eq163} and
integrating [see \ref{appnSur} for the details] we obtain
the leading order $t$-dependence of $Q_{\pm}(t,a)$ as
\begin{equation}
Q_{\pm}(t,a)\sim g_{\pm}(v_0,\alpha,\gamma,a)~ t^{-3/2},
\label{sxtscaleAleAC1}
\end{equation}
where
\begin{equation}
g_{+}(v_0,\alpha,\gamma,a)= 
\frac{a v_0^{3/2} (v_0^2-\alpha^2 )^{5/4}\left(v_0+\sqrt{v_0^2-\alpha ^2}\right)
   \,\,\exp\left[-\frac{a \alpha  \gamma }{v_0
   \sqrt{v_0^2-\alpha ^2}}\right]}
   {\sqrt{2\pi  \gamma}
   \alpha^2 \left[(v_0+\alpha)
   (v_0-\alpha )^3+4 a \alpha  \gamma  \left(a
   \alpha  \gamma
   -(v_0-\alpha )
   \sqrt{v_0^2-\alpha
   ^2}\right)\right]},
\label{sxtscaleAleACCoeff}
\end{equation}
and
\begin{align}
\nonumber
g_{-}&(v_0,\alpha,\gamma,a) = \frac{1}{\sqrt{2\pi} \alpha^2}\left(\frac{v_0}{\gamma }\right)^{3/2} \exp\left[-\frac{a \alpha  \gamma}{v_0 \sqrt{v_0^2-\alpha ^2}}\right]\\
%
&\frac{(v_0-\alpha ) (v_0^2-\alpha^2)^{1/4}
   \left(v_0+\sqrt{v_0^2-\alpha^2}\right)
   \left(v_0^2-\alpha ^2+a \gamma
   \sqrt{v_0^2-\alpha^2}\right)}{\left[2 v_0 \left(\alpha ^3-2 a \alpha  \gamma  \sqrt{v_0^2-\alpha^2}\right)+\alpha ^2 \left(4 a \gamma  \left(a \gamma
   +\sqrt{v_0^2-\alpha^2}\right)-\alpha
   ^2\right)+v_0^4-2 \alpha  v_0^3\right]}.
\label{sxtscaleAleACCoeffneg}
\end{align}
Note that although in {\bf Case-2} ($a=a_c$) and {\bf Case-3}
($a<a_c$) the decay rates of the survival probability are dominated
by the branch point $s_b^+$, the leading order time-dependence of
$Q(t,a)$ are different in these two cases, as the leading singularities are different.

In summary, at large time, we have
\begin{equation}
S_{\pm}(t,a|0) \sim
\left\{
\begin{array}{ll}
\displaystyle
R_{\pm}(a)~e^{-|s^*(a)|t}  & \text{for} ~~ a > a_c \\[2mm]
\displaystyle
f_{\pm}(v_0,\alpha,\gamma,a_c)~
t^{-1/2}~e^{-|s_b^+|t} & \text{for} ~~ a = a_c \\[2mm]
\displaystyle
g_{\pm}(v_0,\alpha,\gamma,a)~
t^{-3/2}~e^{-|s_b^+|t} & \text{for} ~~ 0< a < a_c.
\end{array}
\right.
\label{sxtscaleAll}
\end{equation}

\begin{figure}[t]
\centering
\hspace{2cm}
\includegraphics[scale=0.45]{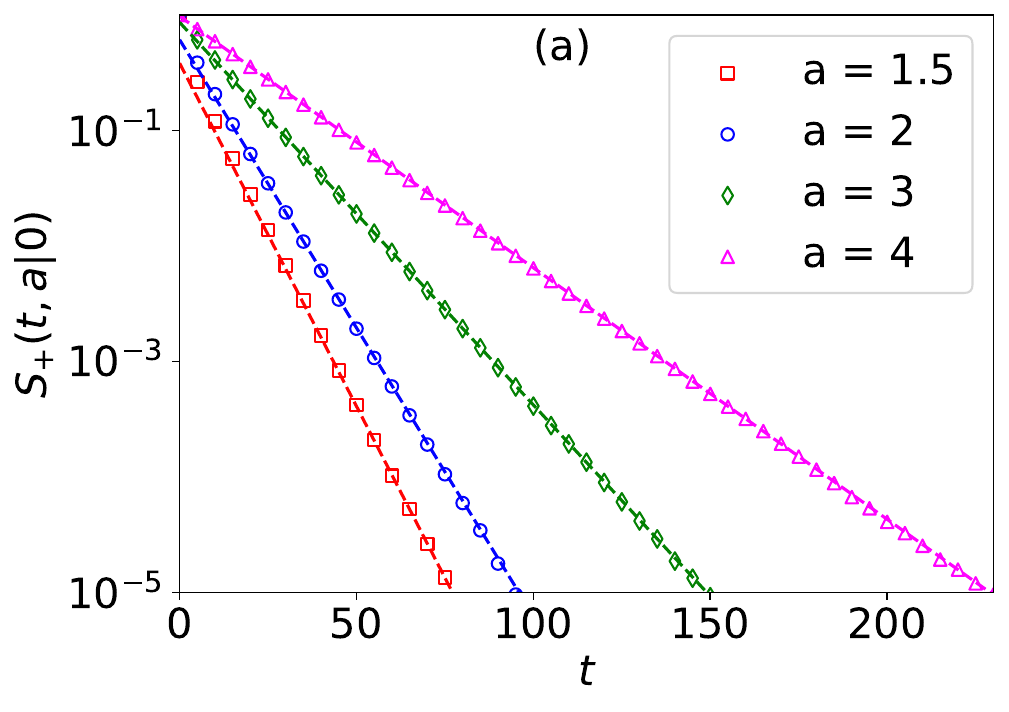}~
\includegraphics[scale=0.45]{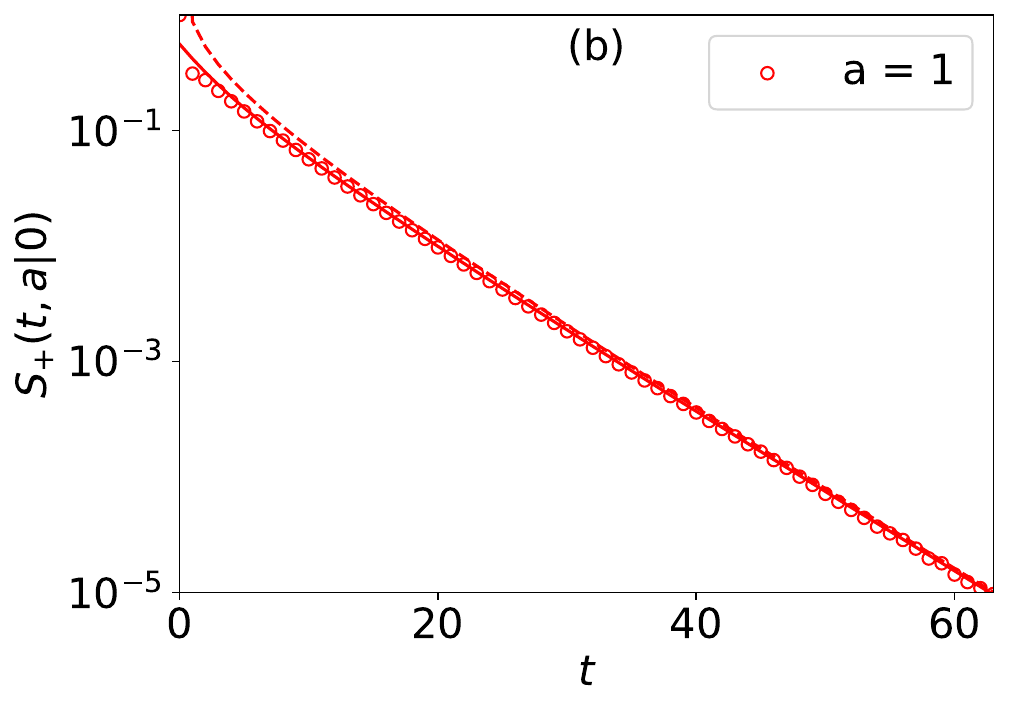} \\
\includegraphics[scale=0.45]{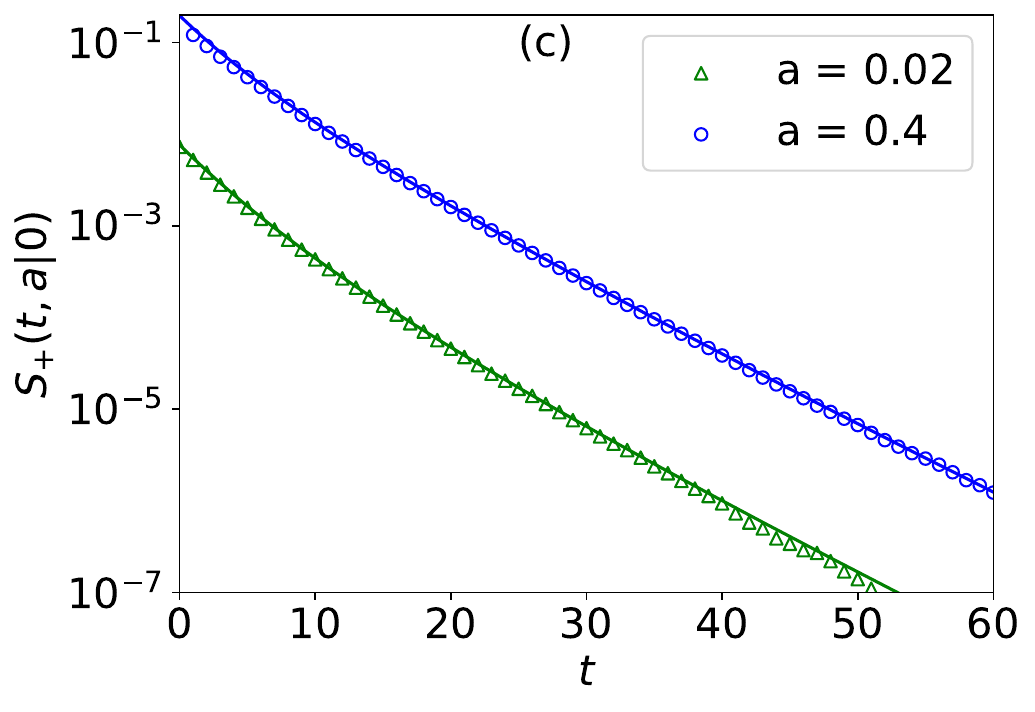}~
\includegraphics[scale=0.45]{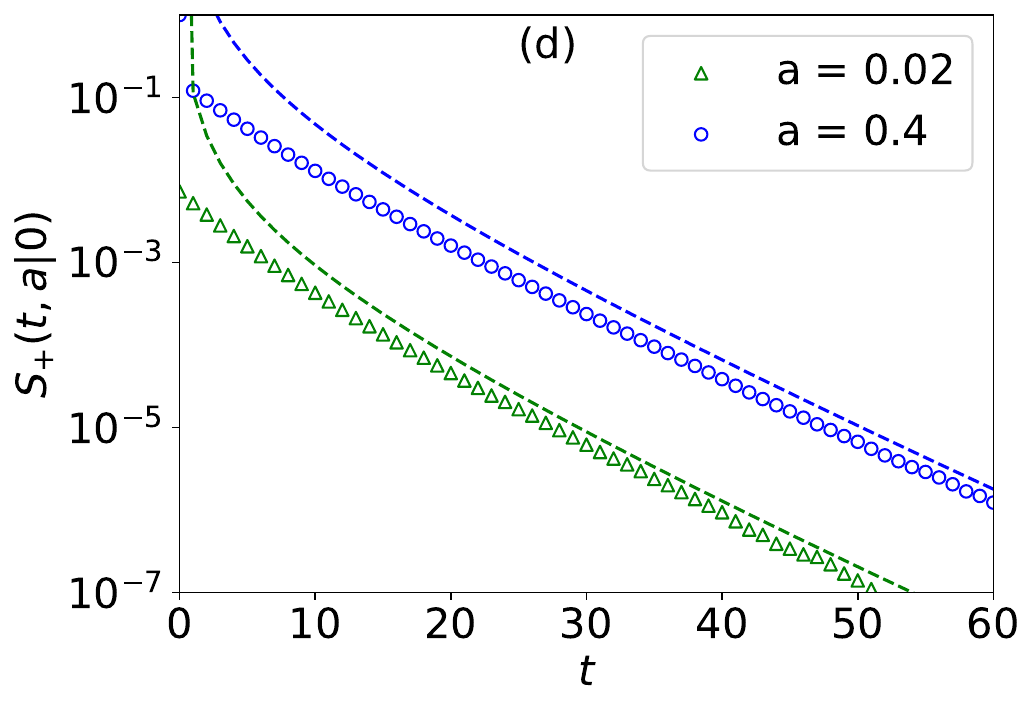}
\caption{Survival probabilities for positive starting velocity $\sigma(0)=+1$.
        (a) Case $a>a_c$: The discrete points are simulated results
        and the corresponding dashed lines are the leading order obtained
        from \eqnref{sxtscaleAll}. (b) Case $a=a_c$: The discrete
        points are simulated results and the corresponding dashed line is
        the leading order obtained from \eqnref{sxtscaleAll}, and the solid
        line is obtained by numerically integrating \eqnref{eq163}.
        (c) Case $a<a_c$: The discrete
        points are simulated results and the corresponding solid
        lines are obtained by numerically integrating \eqnref{eq163}.
        (d) Case $a<a_c$: The discrete
        points are simulated results and the corresponding dashed
        lines are the leading order obtained from \eqnref{sxtscaleAll}.}
\label{figpstat}
\end{figure}

\begin{figure}[t]
\centering
\hspace{2cm}
\includegraphics[scale=0.45]{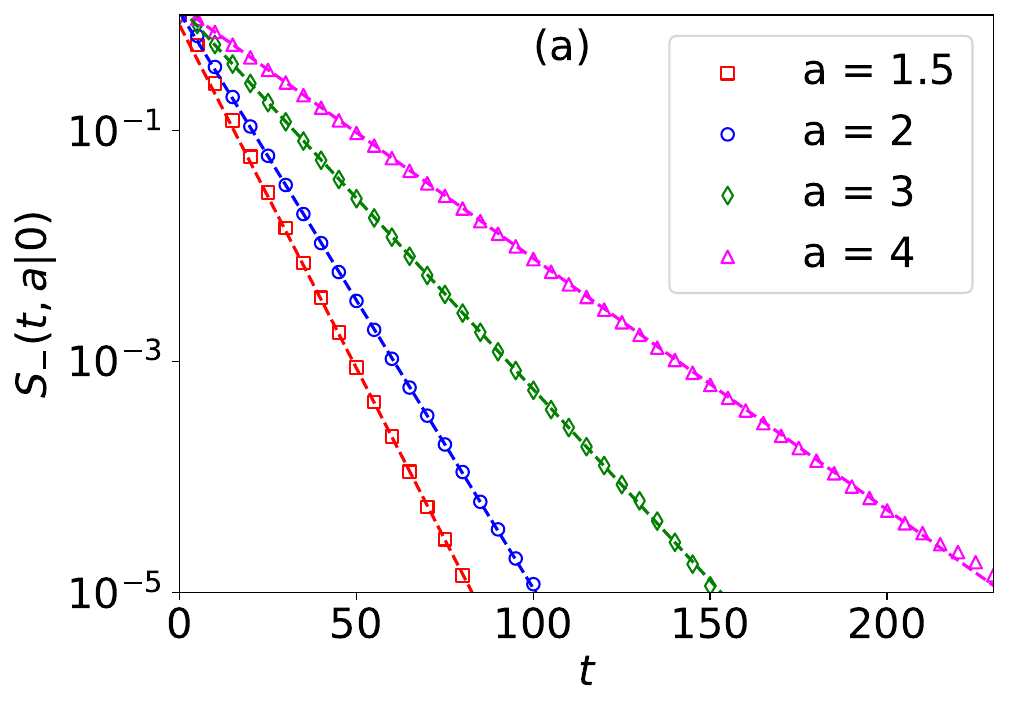}~
\includegraphics[scale=0.45]{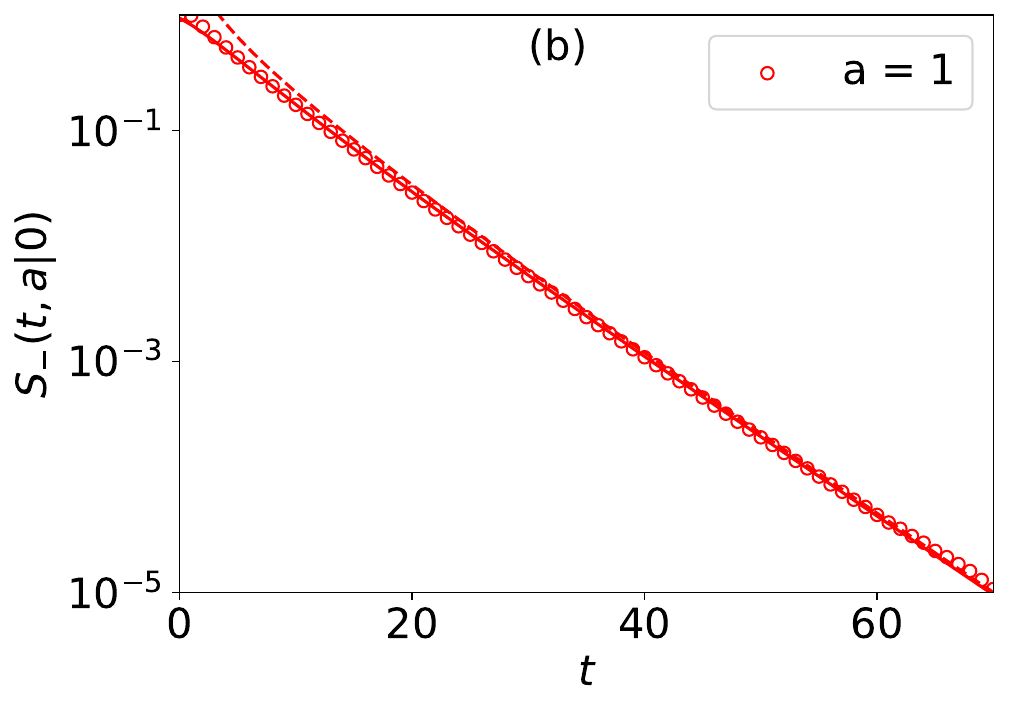} \\
\includegraphics[scale=0.45]{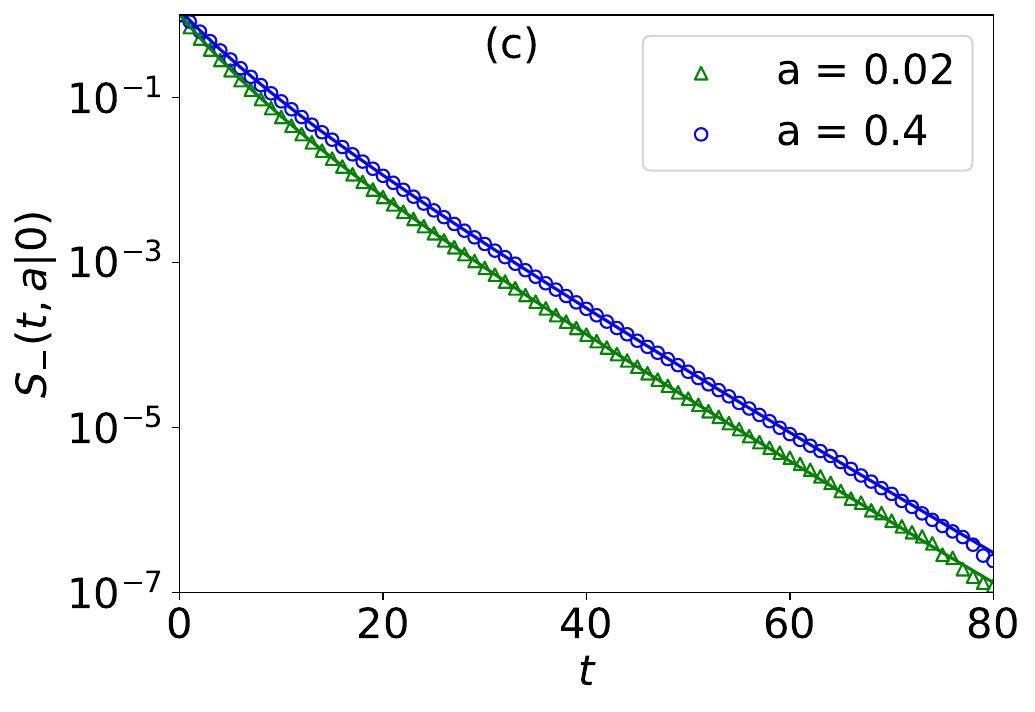}~
\includegraphics[scale=0.45]{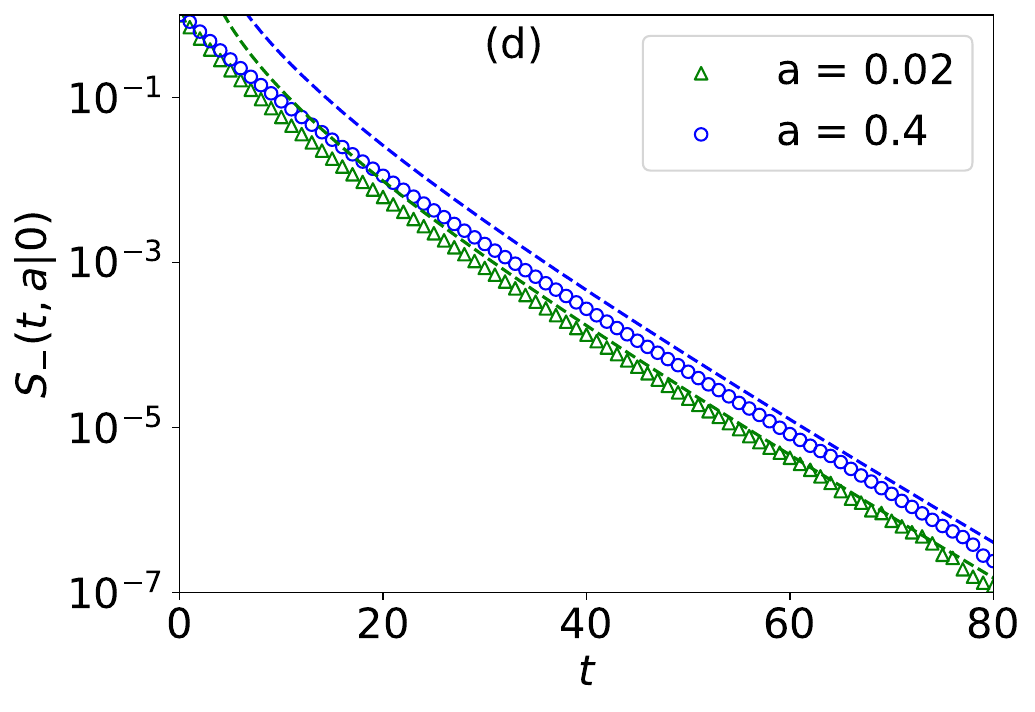}
\caption{Survival probabilities for negative starting velocity $\sigma(0)=-1$.
        (a) Case $a>a_c$: The discrete points are simulated results
        and the corresponding dashed lines are the leading order obtained
        from \eqnref{sxtscaleAll}. (b) Case $a=a_c$: The discrete
        points are simulated results and the corresponding dashed line is
        the leading order obtained from \eqnref{sxtscaleAll}, and the solid
        line is obtained by numerically integrating \eqnref{eq163}.
        (c) Case $a<a_c$: The discrete
        points are simulated results and the corresponding solid
        lines are obtained by numerically integrating \eqnref{eq163}.
        (d) Case $a<a_c$: The discrete
        points are simulated results and the corresponding dashed
        lines are the leading order obtained from \eqnref{sxtscaleAll}.}
\label{fignstat}
\end{figure}

In figures \ref{figpstat} and \ref{fignstat} we plot and compare the survival
probabilities obtained from the stochastic simulations, from numerical integration
of the inverse Laplace Transform in \eqnref{eq163},
and the leading order decay from \eqnref{sxtscaleAll}.
We set  $v_0=5$, $\alpha=4$, and $\gamma=3/8$,
for which we get $a_c=1$. From figures \ref{figpstat}(a) and \ref{fignstat}(a)
we see that, in the regime $a>a_c$, the decay rate of the survival probability
decreases as the distance of the absorbing boundary from the origin increases.
Figures \ref{figpstat}(b) and \ref{fignstat}(b) show the decay rates when the
absorbing boundary is at the critical location $a_c$. Panels
(c) and (d) of figures \ref{figpstat} and \ref{fignstat} plot the decay
rates in the regime $a<a_c$ [see figure captions for details]. Note from the
figures \ref{figpstat}(c) and \ref{fignstat}(c) that although $S_\pm$ have the same
decay rate $\theta(a)$ in the regime $a<a_c$, the prefactor $g_-$ has a smaller variation with respect to its argument $a$, compared to $g_+$.
%
%
\section{Comparison of the decay rate \texorpdfstring{$\theta(a)$}{theta(a)} and the inverse of the
mean first passage time \texorpdfstring{$T_{\pm}(a)$}{Tpm(a)}}
\label{mfptDecayComp}
\begin{figure}[t]
\centering
\hspace{2cm}
\includegraphics[scale=0.45]{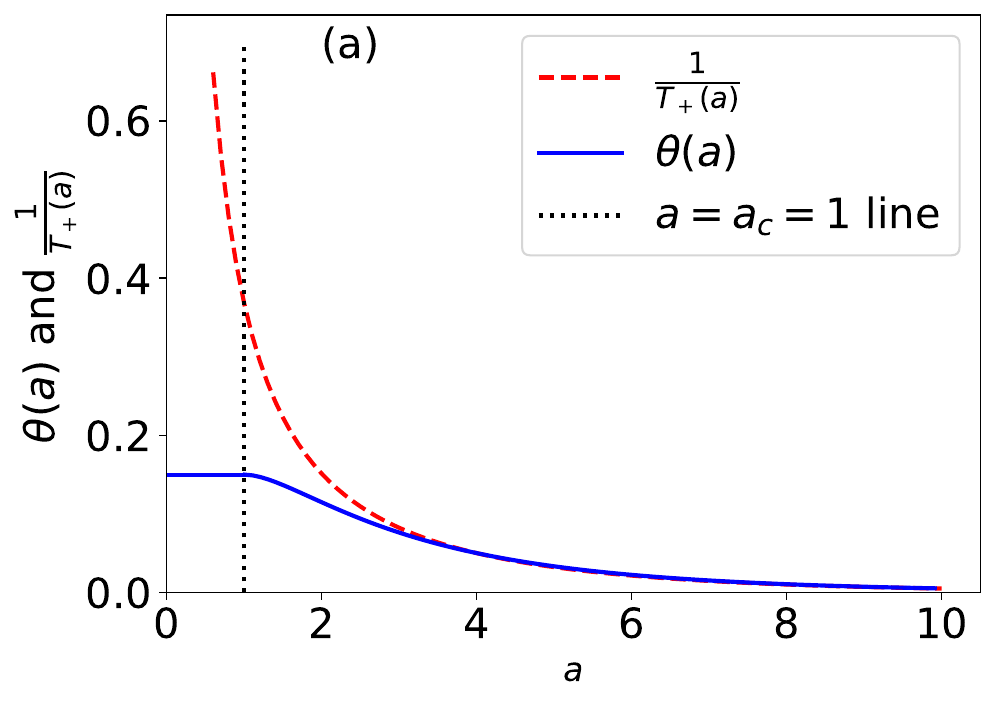}~
\includegraphics[scale=0.45]{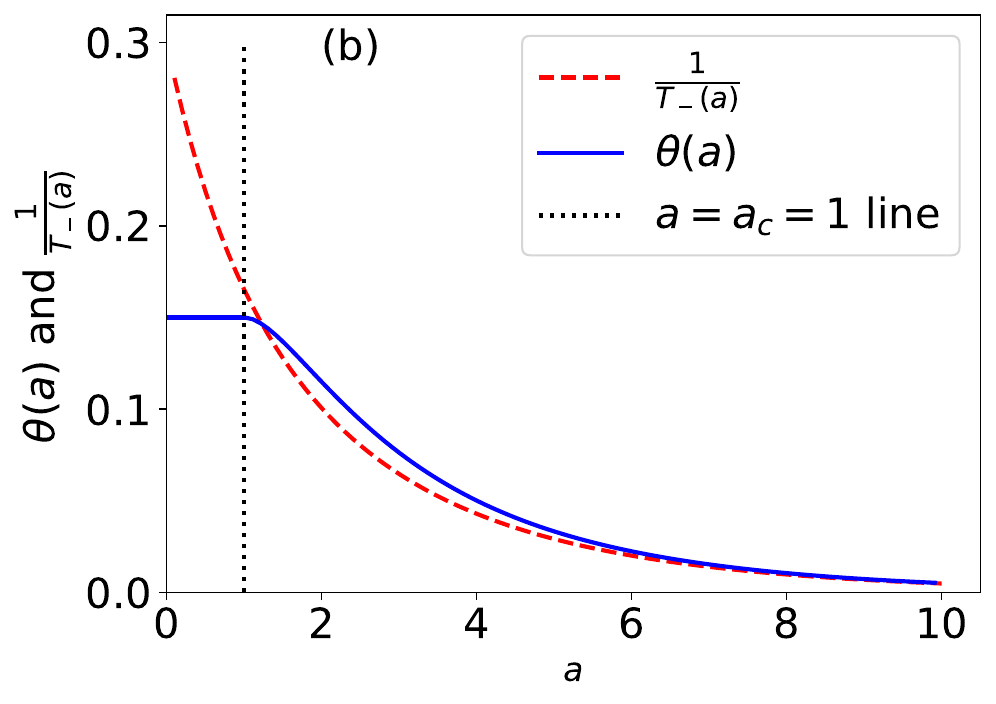}
\caption{Comparison of the inverse of MFPT $T_\pm(a)$, and the decay rate $\theta(a)$. Panels (a) and (b) are for positive and negative starting
        velocities $\sigma(0)=\pm1$, respectively. The red dashed lines are the values of
        $1/T_{\pm}(a)$ obtained from \eqnref{eqmfptApn}.
        The blue solid lines are the values of $\theta(a)$ both
        in the regimes $a>a_c$ and $a\leq a_c$, obtained from \eqnref{thetaA} with numerical evaluation of $p^*(a)$ by solving $\mathbb{D}(p)=0$, where $\mathbb{D}(p)$ is given in \eqnref{eqn168}. The black vertical
        dotted lines denote the critical boundary $a=a_c$.}
\label{figthetaTa}
\end{figure}
Since $S_{\pm}(t,a|0)\approx R_{\pm}(a) e^{-\theta(a)t}$ for  $a\gg a_c$,
[see \eqnref{sxtscaleAll}], the MFPT, $T_{\pm}(a)$, at the absorbing boundary $a$ is
given by
\begin{equation}
T_{\pm}(a)=\int_{0}^{\infty}S_{\pm}(t,a|0)~dt
\approx\frac{R_{\pm}(a)}{\theta(a)}.
\label{mfptSurpRelation}
\end{equation}
Therefore, we have $1/T_{\pm}(a)\propto \theta(a)$ for $a\gg a_c$.
However, for any value of $a$, using the relation
$T_{\pm}(a)=\lim_{s\to0}\tilde{S}_{\pm}(s,a|0)$, and taking $s\to 0$ limit in \eqnref{eq144.aMain} we obtain
\begin{equation}
\hspace{-1.5cm}
T_{+}(a)=-\frac{a}{\alpha}+\frac{v_0 (v_0+\alpha)}{\alpha^2 \gamma}\left(e^{\frac{2 a \alpha  \gamma
   }{v_0^2-\alpha ^2}}-1\right) \quad\text{and}\quad
   T_-(a)=T_+(a)+\frac{v_0}{\alpha\gamma}.
\label{eqmfptApn}
\end{equation}
The MFPT has a non-monotonic behavior as function of $\alpha$. It diverges as $1/\alpha$ as $\alpha\to0$. On the other hand, $T_\pm(a)\sim e^{2 a \alpha \gamma/(v_0^2-\alpha ^2)}$ diverges exponentially as $\alpha\to v_0^-$. There is an optimal value of $\alpha$ where the MFPTs attain minima.
We note that in the diffusive limit, $v_0\to\infty$, $\gamma\to\infty$, keeping $v_0^2/\gamma\to2D$ finite, the above expressions reduce to the passive case
\begin{equation}
T_{+}(a)=T_-(a)=-\frac{a}{\alpha}+\frac{2D}{\alpha^2}
\left(e^{a \alpha / D}-1\right),
\label{eqmfptApnDiffus}
\end{equation}
as discussed in~\cite{PhysRevLett.125.200601}.

In \figref{figthetaTa} we compare $1/T_{\pm}(a)$, calculated using
\eqnref{eqmfptApn}, with the dominant
decay rate $\theta(a)$ in both the regimes $a>a_c$ ($|s^*|$ from
\eqnref{sstarPstar}) and $a\leq a_c$ ($|s_\mathrm{b}^+|$ from \eqnref{sbpm}),
i.e,
\begin{equation}
\theta(a)=
\left\{
\begin{array}{ll}
\displaystyle
\gamma-\frac{1}{v_0}\sqrt{\left[p^{*}(a)\right]^{2}+\gamma^2(v_0^2-\alpha^2)}
& \mbox{if } a > a_c \\ \\
\gamma\left(1-\sqrt{1-(\alpha/v_0)^2}\right)
& \mbox{if } a \leq a_c~, \\
\end{array}
\right.
\label{thetaA}
\end{equation}
where $p^*(a)$ is the root of the equation $\mathbb{D}(p)=0$ for $a > a_c$, where $\mathbb{D}(p)$ is given in \eqnref{eqn168}.
Note that in the regime $a\leq a_c$ although
$\theta(a)$ freezes to a constant value, the inverse of MFPT,
$1/T_{\pm}(a)$, keeps on increasing as $a\to 0$. However, for large $a$
in the regime $a\gg a_c$, $\theta(a)$ and $1/T_{\pm}(a)$ have similar
values. \\ \\
\subsection{Estimation of \texorpdfstring{$1/T_{\pm}(a)$}{1/Tpm(a)} and \texorpdfstring{$\theta(a)$}{theta(a)} for \texorpdfstring{$a\gg a_c$}{a>>ac}:}
In \figref{figthetaTa} we plotted the value of $\theta(a)$, for the regime
$a>a_c$, from \eqnref{thetaA}, with the values of $p^*$ obtained
by solving $\mathbb{D}(p)=0$, numerically, where $\mathbb{D}(p)$ is given by
\eqnref{eqn168}. In this section we derive an analytical estimate
of $\theta(a)$ for the regime $a\gg a_c$ and compare its value
with the corresponding value of $1/T_\pm(a)$. First, we note that
for large $a$, the escape rates, given by
\eqnref{eqmfptApn}, reduce to
\begin{equation}
\frac{1}{T_{+}(a)}\approx \frac{1}{T_{-}(a)} \approx \frac{\alpha^2\gamma}{v_0^2}
\left(1+\frac{\alpha}{v_0}\right)^{-1}
e^{-2 a \alpha \gamma /(v_0^2-\alpha ^2)}.
\label{mfptApnLargeA}
\end{equation}

To obtain an approximate expression for the decay rate $\theta(a)$
in this large $a$ limit, we find an approximate solution
for $s^*(a)$. As $a\to\infty$, from equations \eref{eqn168} and \eref{sstarPstar} we can approximate $p^*=\alpha\gamma-\epsilon_1$ and $s^*(a)=-\epsilon_2$ with $0<\epsilon_1,\epsilon_2\ll 1$. Now,
substituting these values of $s^*$ and $p^*$, in terms of $\epsilon_1$
and $\epsilon_2$, in equation $\mathbb{D}(p,s)=0$ (\eqnref{eqD}),
for large $a$, we obtain
\begin{equation}
\label{ep1ep2Dps0}
\alpha\left[\alpha\gamma-\epsilon_1-v_0\gamma+v_0\epsilon_2)\right]
=v_0\left[-\epsilon_1+\alpha\epsilon_2\right]
e^{2(\alpha\gamma-\epsilon_1)a/(v_0^2-\alpha^2)}.
\end{equation}
Now from \eqnref{sstarPstar} we obtain
\begin{equation}
\epsilon_2
\approx
\frac{\alpha}{v_0^2}~\epsilon_1
\label{ep2}
\end{equation}
Substituting this in \eqnref{ep1ep2Dps0} we get
\begin{equation}
\epsilon_1
\approx \alpha\gamma \left(1+\frac{\alpha}{v_0}\right)^{-1}
e^{-2\alpha\gamma a/(v_0^2-\alpha^2)}.
\label{epsilon1}
\end{equation}
Therefore, $\theta(a)=-s^*(a)=\epsilon_2$ is given by
\begin{equation}
\theta(a)
\approx \frac{\alpha^2\gamma}{v_0^2}\left(1+\frac{\alpha}{v_0}\right)^{-1}
~e^{-2\alpha\gamma a/(v_0^2-\alpha^2)}.
\label{thetaAlargeA}
\end{equation}

To estimate $R_{\pm}(a)$ for large $a$, we substitute $p^*=\alpha\gamma-\epsilon_1$ and $s^*(a)=-\epsilon_2$ in \eqnref{rpmA} with $\epsilon_1,\epsilon_2$ given by \eqnref{epsilon1} and \eqnref{ep2}. This yields
\begin{equation}
R_\pm(a)=1-\frac{v_0}{v_0+\alpha}~e^{-2\alpha\gamma a/(v_0^2-\alpha^2)}
\label{rALargeA}
\end{equation}
Therefore, in the limit $a\gg a_c$, Eqs. \eref{mfptApnLargeA},
\eref{thetaAlargeA}, and \eref{rALargeA} together gives
\begin{equation}
\frac{1}{T_\pm(a)}
\approx \frac{\theta(a)}{R_\pm(a)}
\approx \theta(a)
\approx\frac{\alpha^2\gamma}{v_0^2}\left(1+\frac{\alpha}{v_0}\right)^{-1}
~e^{-2\alpha\gamma a/(v_0^2-\alpha^2)}.
\label{thetaAbyrALargeA}
\end{equation}
In the diffusive limit, $v_0\to\infty$, $\gamma\to\infty$, keeping $v_0^2/\gamma\to2D$ finite, the above expressions reduce to
\begin{equation}
\frac{1}{T_{\pm}(a)} \approx \theta(a)
\approx \frac{\alpha^2}{2D}
~e^{-\alpha a/D}.
\label{thetaAbyrAmfptLargeA}
\end{equation}
Thus in the diffusive limit,
when the absorbing boundary is at a large  distance, or equivalently,
when the height of the potential barrier is large, the escape rate
$1/T_{\pm}(a)$ and the decay rate $\theta(a)$ are same, both having the  Arrhenius form.
\section{The propagator}
\label{RTP-propagator}
In the previous sections we have studied the survival probabilities of the RTP in presence of an absorbing boundary.
Another relevant quantity is the position distribution of the RTP in the presence of the absorbing boundary. This position distribution is not normalized to unity but is expected to decay exponentially with time. In fact, the temporal dependence in the decay of the position distribution 
is exactly the same as that in the survival probability as we will see in this section. 
This because the forward and backward Fokker-Planck equations possess the same eigenvalue spectrum.

In this section, we study the position distribution of the RTP at any time $t$ in the presence 
of an absorbing boundary at $x=a$.
Let $P_\pm(x,t)\, dx$ be the probabilities of finding the particle between the positions $x$ and $x+dx$ at time $t$ with respectively positive and negative velocities. These two probability densities satisfy the forward Fokker-Planck equations
\begin{align}
\label{eqPropagatorDiffEq1}
\frac{\partial P_+}{\partial t} &= - \frac{\partial}{\partial x} \bigl[(-\alpha \mathrm{sgn}(x) + v_0) P_+ \bigr] -\gamma P_+ +\gamma P_-,\\
\label{eqPropagatorDiffEq2}
\frac{\partial P_-}{\partial t} &=  -\frac{\partial}{\partial x} \bigl[(-\alpha \mathrm{sgn}(x) - v_0) P_- \bigr] -\gamma P_- +\gamma P_+.
\end{align}
We consider the initial condition $P_+(x,0)= \beta \delta(x-x_0)$ and $P_-(x,0)=(1-\beta) \delta(x-x_0)$, with $\beta \in [0,1]$. The appropriate boundary condition for the absorbing wall at $a$, is $P_-(a,t)=0$ whereas $P_+(a,t)$ is not specified and related to $P_-(a,t)$ through the above equations. Evidently, $P_\pm(-\infty,t)=0$.
Moreover, integrating the above equation around the origin, we find that the solutions for $x<0$ and $x>0$ satisfy the matching conditions
\begin{align}
(v_0-\alpha)\, P_+(0^+,t) &= (v_0+\alpha)\, P_+(0^-,t),\\
(v_0+\alpha)\, P_-(0^+,t) &= (v_0-\alpha)\, P_-(0^-,t).
\end{align}

\begin{figure}[t]
\includegraphics[angle=0,width=4.7in]{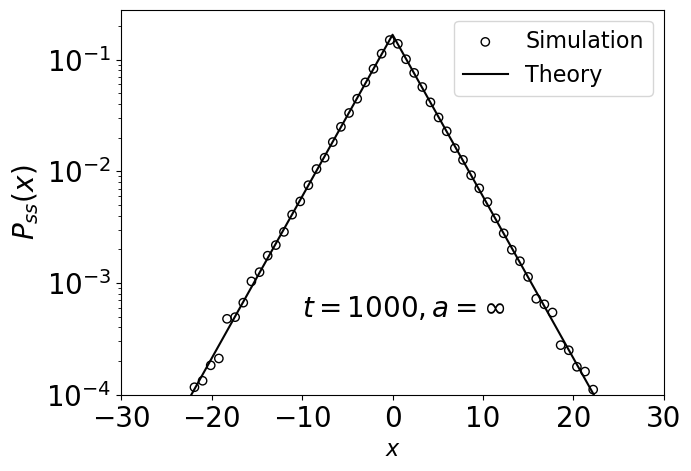}
\caption{Steady state probability density function for
$a\to\infty$.
The discrete points are the simulated data
from the stochastic differential equation (SDE) \eqnref{eq102}, and the solid line
is the theoretical result given in \eqnref{eqpxss}, obtained by solving the corresponding
Fokker-Planck equation. We use $v_0=5, \alpha=4, \gamma=3/8$. The result matches with \cite{dhar2019run}.}
\label{figpospdf}
\end{figure}

The Laplace transforms $\tilde{P}_\pm (x,s)=\int_0^\infty e^{-st}\, P_\pm (x,t)]\, dt$ satisfy
\begin{align}
\label{eqPropagatorLaplaceDiffEq1}
&\frac{\partial}{\partial x} \bigl[(v_0-\alpha \mathrm{sgn}(x)) \tilde{P}_+ \bigr] +(\gamma+s) \tilde{P}_+ -\gamma \tilde{P}_- =\beta\delta(x-x_0),\\
\label{eqPropagatorLaplaceDiffEq2}
&\frac{\partial}{\partial x} \bigl[(v_0+\alpha \mathrm{sgn}(x)) \tilde{P}_- \bigr] -(\gamma+s) \tilde{P}_- +\gamma \tilde{P}_+ =-(1-\beta)\delta(x-x_0).
\end{align}
We have to solve these equations separately in the regions $(-\infty,\min[0,x_0])$, $(\min[0,x_0], \max[0,x_0])$, and $(\max[0,x_0],a)$ and then match the solutions at  $x=0$ and $x=x_0$. Integrating the equations around $x=0$ and $x=x_0$ gives the matching conditions: (1) for $x_0\ne 0$ (strictly)
\begin{align}
\label{eq107.a}
(v_0-\alpha)\,\tilde{P}_+(0^+,s) &= (v_0+\alpha)\,\tilde{P}_+(0^-,s),\\
\label{eq108.a}
(v_0+\alpha)\,\tilde{P}_-(0^+,s) &= (v_0-\alpha)\,\tilde{P}_-(0^-,s),
\end{align}
and (2) for any $x_0$ (including $x_0=0$)
\begin{align}
\label{eq109.a}
\bigl[v_0-\alpha\mathrm{sgn}(x_0^+)\bigr] &\tilde{P}_+(x_0^+,s)  - \bigl[v_0-\alpha\mathrm{sgn}(x_0^-)\bigr] \tilde{P}_+(x_0^-,s)=\beta,\\
\label{eq110.a}
\bigl[v_0+\alpha\mathrm{sgn}(x_0^+)\bigr] &\tilde{P}_-(x_0^+,s)  - \bigl[v_0+\alpha\mathrm{sgn}(x_0^-)\bigr]\tilde{P}_-(x_0^-,s)=-(1-\beta).
\end{align}
Using the boundary conditions and the four matching conditions in Eqs. \eref{eq107.a}, \eref{eq108.a}, \eref{eq109.a}, and \eref{eq110.a} we solve for $\tilde{P}_{\pm}(x,s)$ [see \ref{calsPropagator} for details]. Although the solutions of $\tilde{P}_{\pm}(x,s)$ can be found for any $\beta \in [0,1]$ and $x_0$, here for simplicity, we now consider the case $\beta=1/2$ and $x_0=0$, which corresponds to the initial condition $P_\pm(x,0)= \frac{1}{2}\, \delta(x)$.
For this case,
\begin{align}
\label{pplusxs}
\tilde{P}_+(x,s)&=\frac{1}{4\mathbb{D}}\, e^{-\alpha (\gamma+s)|x|/(v_0^2-\alpha^2)} \\
\nonumber
&\times \left\{
\begin{array}{ll}
a_1^+ e^{p x/(v_0^2-\alpha^2)}
&\text{for}~ x<0,\\[2mm]
b_3^+ e^{-p x/(v_0^2-\alpha^2)}
\left[1 - \left(\frac{v_0(\gamma+s)-p}{v_0(\gamma+s) + p}\right)\,  e^{-2 p (a -x)/(v_0^2-\alpha^2)} \right] &\text{for}~ 0 < x < a,
\end{array}
\right.
\end{align}
and
\begin{align}
\label{pminusxs}
\tilde{P}_-(x,s)&=\frac{1}{4\mathbb{D}}\, e^{-\alpha (\gamma+s)|x|/(v_0^2-\alpha^2)} \\
\nonumber
&\times \left\{
\begin{array}{ll}
a_1^- e^{p x/(v_0^2-\alpha^2)}
&\text{for}~ x<0,\\[2mm]
b_3^- e^{-p x/(v_0^2-\alpha^2)}
\left[1 - e^{-2 p (a -x)/(v_0^2-\alpha^2)} \right] &\text{for}~ 0 < x < a,
\end{array}
\right.
\end{align}
where $p$ and $\mathbb{D}$ are given in \eqnref{eqp} and \eqnref{eqD}, and
\begin{align}
%
\hspace{-2cm}
a_1^- &=
\bigl[p + v_0 (\gamma+s) + \gamma(v_0+\alpha)\bigr]
+\bigl[p - v_0 (\gamma+s) - \gamma(v_0+\alpha)\bigr]e^{-2ap/(v_0^2-\alpha^2)}, \\
a_1^+ &=
\frac{v_0(\gamma+s)-p}{\gamma(v_0+\alpha)}\, a_1^-,\\
b_3^+ &= \bigl[p + v_0 (\gamma+s) + \gamma(v_0+\alpha)\bigr] , \\[2mm]
b_3^- &= \frac{\bigl[v_0(\gamma+s) -p \bigr]}{\gamma(v_0+\alpha)} \, b_3^+.
\label{eq138.a}
\end{align}
Similar expressions for the Laplace transform of the propagator, as
in \eqnref{pplusxs}, \eqnref{pminusxs}, have previously been presented in~\cite{mori2022time}.
Let us first consider the limit $a\to \infty$, where, as discussed in \cite{dhar2019run}, we have
\begin{equation}
\tilde{P}_+(x,s)=\tilde{P}_-(-x,s)=
\frac{e^{-[\alpha (\gamma+s)+p]\,|x|/(v_0^2-\alpha^2)}}{4 v_0 \bigl[p-\alpha (\gamma+s) \bigr]} \bigl[\gamma(v_0+\alpha\,\mathrm{sgn}(x)) + v_0(\gamma+s) + p\,\mathrm{sgn}(x) \bigr].
\end{equation}
Note that $\tilde{P}_\pm(x,s)$ has a pole at $s=0$, i.e., $p=\alpha\gamma$. This gives the steady state distribution for $t\to\infty$,
\begin{equation}
\hspace{-1.5cm}
P_+(x,t\to\infty) = P_-(-x,t\to\infty)=\frac{\alpha\gamma\, e^{-2\alpha\gamma|x|/(v_0^2-\alpha^2)}}{(v_0^2-\alpha^2)}\, \frac{\bigl[v_0 + \alpha \, \mathrm{sgn}(x)\bigr]}{2v_0}.
\end{equation}
Evidently, the unconditional steady-state distribution of $x$
\begin{equation}
\label{eqpxss}
\hspace{-1.5cm}
P(x,t\to\infty)= P_+(x,t\to\infty) + P_-(x,t\to\infty)
= \frac{\alpha\gamma\, e^{-2\alpha\gamma|x|/(v_0^2-\alpha^2)}}{(v_0^2-\alpha^2)},
\end{equation}
which matches with the result obtained in \cite{dhar2019run}, and is
plotted in \figref{figpospdf}.
For any finite value of $a$,
in the limit $t\to\infty$, we have,
\begin{equation}
\label{eqappPosSSa}
P(x,t\to\infty)=P_+(x,t\to\infty) + P_-(x,t\to\infty)
=\lim_{s\to 0}s(\tilde{P}_+(x,s) + \tilde{P}_-(x,s))=0.
\end{equation}
The second equality in \eqnref{eqappPosSSa} is obtained using the Tauberian
theorem of Laplace transform. Equation \eqnref{eqappPosSSa} implies that the
RTP will be absorbed with probability $1$ for any finite position
of the absorbing boundary.

\begin{figure}[t]
\centering
\hspace{2cm}
\includegraphics[scale=0.45]{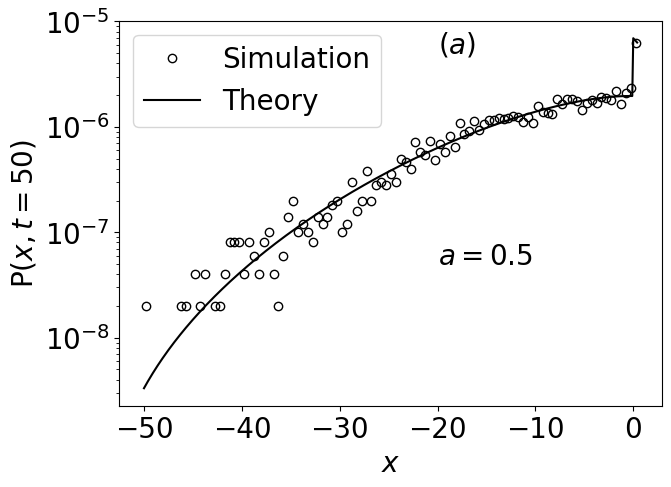}~
\includegraphics[scale=0.45]{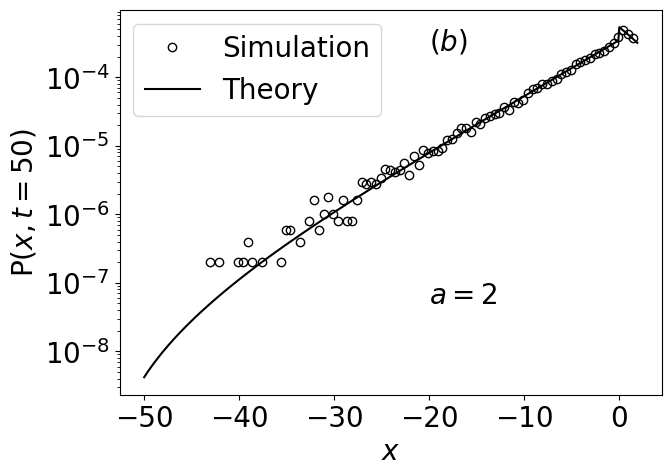}
\caption{Probability density function at $t=50$ for
$v_0=5, \alpha=4, \gamma=3/8, a=0.5$ (panel (a)) and $a=2$ (panel (b)).
The discrete points are the simulated data
from the SDE \eqnref{eq102} and the solid lines
are the theoretical results.}
\label{figpospdfT50a2and0p5}
\end{figure}

To obtain the PDF in the transient state, we can take the inverse Laplace transform of equations \eref{pplusxs} and \eref{pminusxs} as done for the calculation of survival probability in \eqnref{eq163}. In this case we numerically integrate the Bromwich integral
\begin{equation}
\hspace{-1.5cm}
P(x,t)=P_+(x,t)+P_-(x,t)=\frac{1}{2\pi i}\int_{0^+-i\infty}^{0^++i\infty}
\left[\tilde{P}_+(x,s)+\tilde{P}_-(x,s)\right] \, e^{st}\, ds,
\label{eqpxta}
\end{equation}
using the same contour given in \figref{figcontour}.  The analysis is similar to the one carried out in the previous section for the survival probability, and hence, we do not repeat it here. As before, by taking the contributions from the pole (for $a>a_c$) and the branch-point singularities [see \fref{figcontour}], we obtain $P(x,t)$, which we compare with simulations in~\fref{figpospdfT50a2and0p5}.

Essentially, for large $t$,
\begin{equation}
P_{\pm}(x,t) \sim
\left\{
\begin{array}{ll}
\displaystyle
\chi_1^{\pm}(x)~e^{-|s^*(a)|t}  & \text{for} ~~ a > a_c \\[2mm]
\displaystyle
\chi_2^{\pm}(x)~
{t^{-1/2}}~e^{-|s_b^+|t} & \text{for} ~~ a = a_c \\[2mm]
\displaystyle
\chi_3^{\pm}(x)~
{t^{-3/2}}~e^{-|s_b^+|t} & \text{for} ~~ 0< a < a_c,
\end{array}
\right.
\label{pxtscaleAll}
\end{equation}
where the functions $\chi_i^\pm(x)$ with $i=1,2,3$,  can be obtained explicitly.

We end this section by reminding the readers that $\pm$ in the expressions of $P_\pm(x,t)$ refer to the final orientation $\sigma(t)=\pm 1$, whereas in the expressions of the survival probabilities $S_\pm(t,a|x_0)$, they refer to the initial orientation $\sigma(0)=\pm 1$. Therefore, a simple integration of the expressions of $P_\pm(x,t)$ over $x$ does not give $S_\pm(t,a|x_0)$. In order to obtain the survival probabilities, one first needs to obtain the propagators with the initial orientations $\sigma=\pm$, corresponding to $\beta=1$ and $0$, respectively, as discussed in~\ref{survFwdFP}.
\section{Conclusion}
\label{conclusion}
In this paper, we have studied the survival probability and the position distribution
of a one-dimensional run and tumble particle
in the presence of a confining potential $U(x)=\alpha|x|$ and an absorbing
boundary at a finite distance $a$ from the potential minima.
We have analytically
obtained the survival probability, starting from the
origin. At large times $S_\pm(t,a|0)\sim e^{-\theta(a)t}$, and we find a strong dependence of the decay rate $\theta(a)$ on the position of the absorbing boundary.
We find that there is a critical location, $a_c$, such that decay rate freezes to a value $\theta(a_c)$ for $a<a_c$.
Above this critical value, $\theta(a)$ is monotonically decreasing
with increasing value of $a$. We have derived the leading order behaviors
of the survival probabilities in both of these regimes.
Finally, we have compared the escape rates, defined as the inverse of
MFPTs, with the corresponding decay rates. We have shown that in the
regime $a\le a_c$, although the decay rates freeze at a critical value
$\theta(a_c)$, the escape rates keep on increasing with decreasing $a$. We also obtained the exact expression of the MFPT $T_\pm(a)$. For $a\gg a_c$, the two measures of the escape rate, namely $1/T_\pm(a)$ and $\theta(a)$, converge to the corresponding van't Hoff-Arrhenius form. We also obtained the propagator $P_\pm(x,t)$ in the presence of the absorbing boundary at $x=a$. We verified our analytical results with numerical simulations.

There are many possible extensions and open questions which could be
of interest. Our present formulation can be extended to study the effect of randomly switching potentials on the first passage properties of an RTP, as studied in \cite{mercado2020intermittent,santra2021brownian,mercado2022reducing} for Brownian particles. There have been some recent studies on stochastic resetting for a two dimensional RTP \cite{santra2020run}. However, resetting by switching potentials would be very intriguing as they are much easier to reproduce in experiments.
Another interesting question to study is the relaxation to the steady state
and the first passage properties of higher dimensional RTPs under confining potentials. Although some studies have been done in this direction~\cite{fodor2016far,fox1986uniform,haunggi1994colored}, systematic studies of the first passage properties are still scarce in this domain. Finally,
It would be very intriguing to investigate
whether similar freezing transitions exist for other types of active models
such as active Brownian particle, Active Ornstein-Uhlenbeck Particle, and direction-reversing active Brownian particles~\cite{santra2021direction}.
\section*{acknowledgments}
SS thanks A. Kundu and S. N. Majumdar for useful discussions. SKN acknowledges the Leverhulme Trust (RPG-2018-137) for supporting
his time and resources for research in the UK. This work was partially undertaken on ARC3, part of the High Performance Computing facilities at the University of Leeds, UK and the Computational Shared Facility at The University of Manchester, UK.

\appendix
\label{appendix}
\section{Details of calculations of the survival probability in \sref{spMain}}
\label{calsSurv}

In this appendix we give some details for the calculation of the survival probability
given in \sref{spMain}.
To solve for $\tilde{S}_\pm (s,a|x_0)$ from \eqnref{eq105} and \eqnref{eq106}, it is convenient to define
\begin{equation}
\label{eqStildesa}
\tilde{S}_\pm (s,a|x_0)=\frac{1}{s} +\tilde{q}_\pm(s,a|x_0),
\end{equation}
such that $\tilde{q}_\pm(s,a|x_0)$ satisfy a set of coupled first order homogeneous equations
\begin{align}
\label{eq118}
\frac{\partial}{\partial x_0}
\left( {\begin{array}{cc}
   \tilde{q}_+  \\[3mm]
   \tilde{q}_-  \\
 \end{array} } \right)
 =\left( {\begin{array}{cc}
   \frac{\gamma+s}{v_0-\alpha\mathrm{sgn}(x_0)} & \frac{-\gamma}{v_0-\alpha\mathrm{sgn}(x_0)} \\[3mm]
   \frac{\gamma}{v_0+\alpha\mathrm{sgn}(x_0)} & \frac{-(\gamma+s)}{v_0+\alpha\mathrm{sgn}(x_0)}  \\
 \end{array} } \right)
 \left( {\begin{array}{cc}
   \tilde{q}_+  \\[3mm]
   \tilde{q}_-  \\
 \end{array} } \right).
\end{align}
The general solution for $\tilde{q}_\pm$ can be readily written down in terms of the eigenvalues of the $2\times2$ matrix on the right hand side. The eigenvalues are given by
\begin{equation}
\lambda_{\pm}=\frac{\alpha\,\text{sgn}(x_0)(\gamma+s)\pm p}{(v_0^2-\alpha^2)},
\label{eqeigenvalues}
\end{equation}
which in turn gives
\begin{align}
\label{eq108.a2}
\tilde{q}_\pm (s,a|x_0) =  e^{\alpha (\gamma + s) |x_0|/(v_0^2-\alpha^2)}
\left\{
\begin{array}{ll}
A_1^\pm e^{p x_0/(v_0^2-\alpha^2)} &\text{for}~~ x_0<0 \\[2mm]
A_2^\pm e^{p x_0/(v_0^2-\alpha^2)} + B_2^\pm e^{-p x_0/(v_0^2-\alpha^2)} &\text{for}~~ x_0>0
\end{array}
\right.
\end{align}
where $p$ is given by
\begin{equation}
p=\sqrt{v_0^2(\gamma+s)^2-\gamma^2(v_0^2-\alpha^2)},
\label{eqp1}
\end{equation}
and we have already implemented the boundary condition $\tilde{q}_\pm (s,a|-\infty)=0$. To determine the six constants $\{A_{1,2}^\pm$, $B_2^\pm\}$,
we substitute the above solutions (obtained in \eqnref{eq108.a2})
in~\eqnref{eq105} (or \eqnref{eq106}), and obtain
\begin{align}
A_1^- &= A_1^+  \frac{[v_0(\gamma+s)-p]}{\gamma (v_0-\alpha)},\\
A_2^- &= A_2^+\frac{[v_0(\gamma+s)-p]}{\gamma (v_0+\alpha)},\\
B_2^- &= B_2^+  \frac{[v_0(\gamma+s)+ p]}{\gamma (v_0+\alpha)}.
\end{align}
Integrating Eqs.~\eref{eq105} and \eref{eq106} around the origin gives the continuity relations $\tilde{S}_+(s,a|0^+)= \tilde{S}_+(s,a|0^-)$ and $\tilde{S}_-(s,a|0^+)=\tilde{S}_-(s,a|0^-)$. These relations together with the boundary condition $\tilde{S}_+(s,a|a)=0$ determine all the constants. Skipping details,
\begin{align}
A_1^+ &^= \frac{-1}{s\mathbb{D}}p (v_0-\alpha)\, e^{-[p+\alpha(\gamma+s)]a/(v_0^2-\alpha^2)},\\
A_2^+ &^= \frac{-1}{s\mathbb{D}} v_0 [p-\alpha(\gamma+s)]\, e^{-[p+\alpha(\gamma+s)]a/(v_0^2-\alpha^2)},\\
B_2^+ &^= \frac{1}{s\mathbb{D}} \alpha[p-v_0(\gamma+s)]\, e^{-[p+\alpha(\gamma+s)]a/(v_0^2-\alpha^2)},
\end{align}
where $\mathbb{D}$ is given by
\begin{equation}
\mathbb{D}=v_0\bigl[p-\alpha (\gamma+s) \bigr]-\alpha \bigl[p-v_0(\gamma+s) \bigr] e^{-2 p a/(v_0^2-\alpha^2)}.
 \label{eqD1}
\end{equation}
It is easily seen that substituting these constants in \eqnref{eq108.a2} and using \eqnref{eqStildesa} gives \eref{eq144.aMain}.

\section{Details of calculations of the propagator in  \sref{RTP-propagator}}
\label{calsPropagator}
In this section we provide details of the solution for the propagators $\tilde{P}_{\pm}(x,s)$ in Laplace domain using the boundary conditions described after \eqnref{eqPropagatorDiffEq2} and the four matching conditions in Eqs. \eref{eq107.a}, \eref{eq108.a}, \eref{eq109.a}, and \eref{eq110.a}.
Away from the points $x=0$ and $x=x_0$, we get the homogeneous equations (taking Laplace transform of \eqnref{eqPropagatorDiffEq1} and \eqnref{eqPropagatorDiffEq2})
\begin{align}
\label{eq111.a}
\bigl[v_0-\alpha \mathrm{sgn}(x)\bigr]\tilde{P}'_+  +(\gamma+s) \tilde{P}_+ -\gamma \tilde{P}_- =0,\\
\label{eq112.a}
\bigl[v_0+\alpha \mathrm{sgn}(x) \bigr] \tilde{P}'_- -(\gamma+s) \tilde{P}_- +\gamma \tilde{P}_+ =0,
\end{align}
which satisfy the general solutions [see the discussion after \eqnref{eq118}]
\begin{align}
\label{eq113.a}
\tilde{P}_+(x,s)&=A_i^+ e^{m_+ x} + B_i^+ e^{m_- x},\\
\label{eq114.a}
\tilde{P}_-(x,s)&=A_i^-e^{m_+ x} + B_i^-e^{m_- x},
\end{align}
where the subscript $i=1,2,3$ represents the three regions $(-\infty,\min[0,x_0])$, $(\min[0,x_0], \max[0,x_0])$, and $(\max[0,x_0],a)$ respectively.
Not all of the four constants $A_i^\pm$ and $B_i^\pm$, in each region,  are independent of each other and relation between them can be found by substituting the above solutions in \eqnref{eq111.a} and then setting the coefficients of $e^{m_+ x}$ and $e^{m_- x}$ to zero respectively:
\begin{align}
\label{eq115.a}
A_i^-&=\frac{A_i^+}{\gamma}\left[(\gamma+s) + m_+\bigl(v_0-\alpha\mathrm{sgn}(x)\bigr) \right],\\
\label{eq116.a}
B_i^-&=\frac{B_i^+}{\gamma}\left[(\gamma+s) + m_-\bigl(v_0-\alpha\mathrm{sgn}(x)\bigr) \right].
\end{align}
Substituting the solutions \eqnref{eq113.a} and \eqnref{eq114.a} in \eqnref{eq112.a} and using Eqs.\eref{eq115.a} and \eref{eq116.a} determines $m_\pm$ as the two solutions of
\begin{equation}
\bigl(v_0^2-\alpha^2\bigr) \,m_\pm^2 + 2\alpha (\gamma +s)\,\mathrm{sgn}(x) \,m_\pm
- \left[(\gamma+s)^2-\gamma^2 \right] =0.
\end{equation}
Therefore,
\begin{equation}
m_\pm =\frac{-\alpha(\gamma+s)\, \mathrm{sgn}(x) \pm p}{v_0^2-\alpha^2},
\end{equation}
where
\begin{equation}
p=\sqrt{v_0^2(\gamma+s)^2-\gamma^2(v_0^2-\alpha^2)}~~.
\label{eq119.a}
\end{equation}
Using $m_\pm$, Eqs.~\eref{eq115.a} and \eref{eq116.a} can be simplified to
\begin{align}
\label{eq122.a}
A_i^-&=A_i^+\,\frac{v_0(\gamma+s) +p}{\gamma \left[v_0+\alpha\mathrm{sgn}(x) \right]},\\
\label{eq123.a}
B_i^-&=B_i^+\,\frac{v_0(\gamma+s) -p}{\gamma \left[v_0+\alpha\mathrm{sgn}(x) \right]}.
\end{align}
The boundary condition $\tilde{P}_-(a,s)=0$  yields
\begin{equation}
A_3^- = -B_3^-\,e^{-2 p a/(v_0^2-\alpha^2)},
\end{equation}
which according to Eqs. \eref{eq122.a} and \eref{eq123.a} also
implies
\begin{equation}
A_3^+ = - B_3^+ \left[\frac{v_0(\gamma+s) -p}{v_0(\gamma+s)+p}\right] \,e^{-2 p a/(v_0^2-\alpha^2)}.
\end{equation}
On the other hand, the boundary condition $\tilde{P}_\pm(-\infty,s)=0$ gives $B_1^\pm=0$. Therefore, the solutions can be expressed as
\begin{align}
\label{eq126.a}
&\tilde{P}_+(x,s)=e^{-\alpha (\gamma+s)|x|/(v_0^2-\alpha^2)} \\
\nonumber
&\times \left\{
\begin{array}{ll}
A_1^+ e^{p x/(v_0^2-\alpha^2)} &\text{if}~ x < \min(0,x_0)\\
A_2^+ e^{p x/(v_0^2-\alpha^2)} + B_2^+ e^{-p x/(v_0^2-\alpha^2)}
&\text{if} \min(0,x_0) < x < \max(0,x_0)\\
B_3^+ \left[e^{-p x/(v_0^2-\alpha^2)} - \left(\frac{v_0(\gamma+s)-p}{v_0(\gamma+s) + p}\right)\,  e^{-p (2a -x)/(v_0^2-\alpha^2)} \right]
&\text{if} \max(0,x_0) < x < a
\end{array}
\right.
\end{align}
and
\begin{align}
\label{eq127.a}
\tilde{P}_-(x,s)&=e^{-\alpha (\gamma+s)|x|/(v_0^2-\alpha^2)} \\
\nonumber
&\times\left\{
\begin{array}{ll}
A_1^- e^{p x/(v_0^2-\alpha^2)} &\text{if}~ x < \min(0,x_0)\\
A_2^- e^{p x/(v_0^2-\alpha^2)} + B_2^- e^{-p x/(v_0^2-\alpha^2)}
&\text{if}~ \min(0,x_0) < x < \max(0,x_0)\\
B_3^-\left[e^{-p x/(v_0^2-\alpha^2)} -   e^{-p (2a -x)/(v_0^2-\alpha^2)} \right] &\text{if}~ \max(0,x_0) < x < a
\end{array}
\right.
\end{align}
where $A_1^-$, $A_2^-$, $B_2^-$ and $B_3^-$ are related to $A_1^+$, $A_2^+$, $B_2^+$, and $B_3^+$ respectively by
\begin{align}
\label{eq:const_A1}
A_1^-&=A_1^+\,\frac{[v_0(\gamma+s)+p]}{\gamma (v_0-\alpha)},\\
A_2^-&=A_2^+ \,\frac{[v_0(\gamma+s)+p]}{\gamma [v_0+\alpha\mathrm{sgn}(x_0)]},\\
B_2^-&=B_2^+ \,\frac{[v_0(\gamma+s)-p]}{\gamma[v_0+\alpha\mathrm{sgn}(x_0)]},\\
B_3^- &= B_3^+ \, \frac{[v_0(\gamma+s)-p]}{\gamma (v_0+\alpha)}.
\label{eq:const_B3}
\end{align}
The remaining four independent constants $A_1^+$, $A_2^+$, $B_2^+$ and $B_3^+$ are determined by the four matching conditions in Eqs. \eref{eq107.a}, \eref{eq108.a}, \eref{eq109.a}, and \eref{eq110.a}.

\section{Survival probability by integrating the propagator}
\label{survFwdFP}
Let $S_\pm (t,a|x_0)$ be the survival probability starting at position $x_0$, with positive ($\sigma(0)=1$, which corresponds to $\beta=1$ in \sref{RTP-propagator}) and negative ($\sigma(0)=-1$, which corresponds to $\beta=0$ in \sref{RTP-propagator}) velocity respectively. Therefore,
\begin{align}
S_+(t,a|x_0)&= \int_{-\infty}^a \bigl[P_+(x,t) + P_-(x,t)\bigr]\, dx \quad\text{with}~\beta=1,\\
S_-(t,a|x_0)&= \int_{-\infty}^a \bigl[P_+(x,t) + P_-(x,t)\bigr]\, dx \quad\text{with}~\beta=0,
\end{align}
where the Laplace transform of $P_\pm(x,t)$ are given in \eref{eq126.a} and \eref{eq127.a}. We can eliminate four of the constants in terms of the others using \eref{eq:const_A1}--\eref{eq:const_B3}. We determine the remaining four constants using the conditions  \eref{eq107.a}--\eref{eq110.a} for arbitrary $\beta$. This determines the Laplace transforms $\tilde{P}_\pm(x,s)$ completely for arbitrary $\beta$. 
Setting $\beta=1$ and $0$ respectively, and subsequently integrating $\tilde{P}_+(x,s) + \tilde{P}_-(x,s)$ over $x$, yields the expressions for the Laplace transforms $\tilde{S}_\pm (s,a|x_0) = \int_0^\infty S_\pm (t,a|x_0)\, e^{-s t}\, dt$, as
\begin{equation}
\label{eq144.a}
\tilde{S}_\pm(s,a|x_0)=\frac{1}{s}\Bigl[1-\tilde{F}_\pm (s,a|x_0) \Bigr],
\end{equation}
with
\begin{equation}
\tilde{F}_\pm(s,a|x_0)= e^{-[\alpha (\gamma+s)(a-|x_0|)+pa]/(v_0^2-\alpha^2)}\, \tilde{\psi}_\pm(s|x_0).
\end{equation}
The functions $\tilde{\psi}_\pm(s|x_0)$ in the above expression are given separately for  the two cases $0 < x_0 \le a$ and $x_0<0$. 

For $0 < x_0 \le a$, we have
\begin{align}
\tilde{\psi}_+(s|x_0)&=\frac{1}{\mathbb{D}}\,
\Bigl\{
v_0\bigl[p-\alpha (\gamma+s) \bigr]e^{px_0/(v_0^2-\alpha^2)}-\alpha \bigl[p-v_0(\gamma+s) \bigr] \,e^{- p x_0/(v_0^2-\alpha^2)}
\Bigr\}, \\
\nonumber
\tilde{\psi}_-(s|x_0)&=\frac{1}{\mathbb{D}}\,
\left[\frac{v_0(\gamma+s)-p}{\gamma(v_0+\alpha)}\right] \\
&\Bigl\{
v_0\bigl[p-\alpha (\gamma+s) \bigr]\,e^{px_0/(v_0^2-\alpha^2)}+\alpha \bigl[p+v_0(\gamma+s) \bigr] e^{- p x_0/(v_0^2-\alpha^2)}
\Bigr\},
\end{align}
where $\mathbb{D}$ is given by \eref{eqD}.
It is easily checked that $\tilde{F}_+(s,a|a)=1$, and consequently,
$\tilde{S}_+(s,a|a)=0$.

For $x_0<0$ we have
\begin{align}
\label{eq:psiPForward}
\tilde{\psi}_+(s|x_0)&=\frac{p}{\mathbb{D}}\,(v_0-\alpha)\, e^{px_0/(v_0^2-\alpha^2)}\\
\label{eq:psiNForward}
\tilde{\psi}_-(s|x_0)&=\frac{p}{\gamma\mathbb{D}}\,\big[v_0(\gamma+s)-p\bigr]\, e^{px_0/(v_0^2-\alpha^2)}.
\end{align}

It is easily checked that $\tilde{F}_\pm(s,a|x_0) \to 0$ for both $x_0\to -\infty$ (for a given $a$) and $a\to \infty$ (for a given $x_0$), and consequently,  $\tilde{S}_\pm(s,a|x_0)\to1/s$ i.e.,  $S_\pm(t,a|x_0)\to 1$. On the other hand, for any finite $a$, we have $\tilde{F}_\pm(s\to 0,a|x_0) = 1+ O(s)$, and hence, $s=0$ is no longer a pole. We also note the continuity of the $\tilde{F}_\pm$ across $x_0=0$, i.e., $\tilde{F}_+(s,a|0^+)=\tilde{F}_+(s,a|0^-)$ and $\tilde{F}_-(s,a|0^+)=\tilde{F}_-(s,a|0^-)$.

Thus, we have verified that the Laplace transform $\tilde{S}_\pm(s, a|x_0)$ of the survival probabilities, 
in \eref{eq144.a}--\eref{eq:psiNForward} obtained here by integrating the propagator are exactly the same as \eref{eq144.aMain}--\eref{eq:psiNBackward}  obtained earlier in \secref{spMain}
 directly from the Backward Fokker-Planck.
\section{Contour integration for evaluating the survival probability in
\texorpdfstring{$(x,t)$}{(x,t)} space}
\label{appnSur}

The survival probability $S_{\pm}(t,a|x_0)$, for $x_0=0$, is given by the Bromwich integral
\begin{equation}
S_{\pm}(t,a|0)=\frac{1}{2\pi i}\int_{0^+-i\infty}^{0^++i\infty}
S_{\pm}(s,a|0)~e^{st}ds.
\label{appEqsxt1}
\end{equation}
\begin{figure}[t]
\begin{center}
\includegraphics[angle=0,width=2.5in]{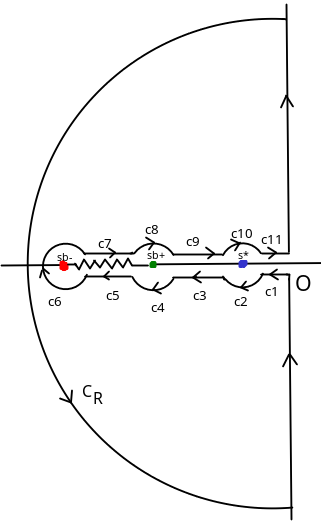}
\caption{Construction of the Bromwich contour. Red, green, and blue points
are the schematic positions of the branch points $s_b^-$, $s_b^+$, and the
pole $s^*$, respectively.}
\label{figcontour}
\end{center}
\end{figure}
Considering the integral along the  contour, as drawn in
\figref{figcontour}.  Using Cauchy-Goursat theorem,  we obtain
\begin{align}
\nonumber
\frac{1}{2\pi i}\int_{0^+-i\infty}^{0^++i\infty} & S_{\pm}(s,a|0)~e^{st}ds= \\
& -\frac{1}{2\pi i}\sum_{j=1}^{11} \int_{C_{j}} S_{\pm}(s,a|0)~e^{st}ds
-\frac{1}{2\pi i}\int_{C_{R}} S_{\pm}(s,a|0)~e^{st}ds.
\label{appEqsxt2}
\end{align}
It can be shown that the integrals on the contours $C_1$ and $C_3$
are canceled by the integrals along the contours $C_{11}$ and
$C_9$, respectively. The integral around the pole $s^*$
(on the contours $C_2$ and $C_{10}$) is given
by the residue of $S_{\pm}(s,a|0)$ at $s^*$. The integrals around the
branch points $s_b^+$ (on the contours $C_4$ and $C_8$) and $s_b^-$
(on the contour $C_6$) and the integral along the
circular arc $C_R$ tend to zero. Therefore, as there is no pole for
$a\leq a_c$, we obtain
\begin{align}
\nonumber
\frac{1}{2\pi i}\int_{-i\infty}^{+i\infty} & S_{\pm}(s,a|0)~e^{st}ds \\
\nonumber
=& -\frac{1}{2\pi i}\int_{C_{5}} S_{\pm}(s,a|0)~e^{st}ds
-\frac{1}{2\pi i}\int_{C_{7}} S_{\pm}(s,a|0)~e^{st}ds \\
=& -\frac{1}{2\pi i}\int_{s_b^+}^{s_b^-} S_{\pm5}(s,a|0)~e^{st}ds
-\frac{1}{2\pi i}\int_{s_b^-}^{s_b^+} S_{\pm7}(s,a|0)~e^{st}ds,
\label{appEqsxt3}
\end{align}
where $S_{\pm5}(s,a|0)$ and $S_{\pm7}(s,a|0)$ are the values of
$S_{\pm}(s,a|0)$ on the contours $C_5$ and $C_7$, respectively.
For $a > a_c$, as there is a pole $s^*$ of the integrand
in the interval $(s_b^+,0)$, we
get
\begin{align}
\nonumber
\frac{1}{2\pi i} & \int_{0^+-i\infty}^{0^++i\infty} S_{\pm}(s,a|0)~e^{st}ds
= -\frac{1}{2\pi i}\int_{C_{5}} S_{\pm}(s,a|0)~e^{st}ds
-\frac{1}{2\pi i}\int_{C_{7}} S_{\pm}(s,a|0)~e^{st}ds \\
&=-\frac{1}{2\pi i}\int_{s_b^+}^{s_b^-} S_{\pm5}(s,a|0)~e^{st}ds
-\frac{1}{2\pi i}\int_{s_b^-}^{s_b^+} S_{\pm7}(s,a|0)~e^{st}ds
+R_{\pm}~e^{-|s^*|t},
\label{appEqsxt3r}
\end{align}
When the particle starts with a positive initial velocity,
for $j\in\{5, 7\}$, we obtain
\begin{equation}
\hspace{-1.5cm}
S_{+j}(s,a|0)= \frac{1}{s} \left( 1+\frac{q_j (\alpha -v_0) e^{a (q_j-\alpha  (\gamma
   +s))/(v_0^2-\alpha^2 )}}{v_0 e^{2 a
   q_j/(v_0^2-\alpha ^2)} (q_j-\alpha  \left| s+\gamma \right| )+\alpha
   v_0 \left| s+\gamma \right| -\alpha q_j}\right),
\label{apps0sc5}
\end{equation}
and for negative initial velocity we obtain
\begin{equation}
\hspace{-1.5cm}
S_{-j}(s,a|0)= \frac{1}{s} \left( 1+\frac{\frac{q_j}{\gamma} (q_j-(v_0\gamma+s)) e^{a (q_j-\alpha  (\gamma
   +s))/(v_0^2-\alpha^2 )}}{v_0 e^{2 a
   q_j/(v_0^2-\alpha ^2)} (q_j-\alpha  \left| s+\gamma \right| )+\alpha
   v_0 \left| s+\gamma \right| -\alpha q_j}\right),
\label{apps0sc5c7neg}
\end{equation}
where $q_5=-i\sqrt{-\alpha^2 \gamma ^2-s^2 v_0^2-2\gamma sv_0^2}$,
and $q_7=i\sqrt{-\alpha^2 \gamma ^2-s^2 v_0^2-2\gamma sv_0^2}$.

%
\section*{References}
\bibliographystyle{iopart-num}
\bibliography{rtmodXiop}
\end{document}